%% file: main.tex
\documentclass[11pt, a4paper]{report}

\usepackage[utf8]{inputenc}
\usepackage[T1]{fontenc}
\usepackage{lmodern}

\usepackage{textcomp}
\usepackage{gensymb}
\usepackage{blindtext}
\usepackage{enumitem}
\setlist{parsep=0pt,listparindent=\parindent}
\usepackage{amsmath}
\usepackage{breqn}
\usepackage{float}
\usepackage{caption}
\usepackage[section]{placeins}
\usepackage{lipsum}  
\usepackage{graphicx}
\usepackage{caption}
\extrafloats{100}
\usepackage{cite}
\usepackage{listings}
\usepackage[dvipsnames]{xcolor}
\usepackage{listings}
\usepackage{amsfonts}
\usepackage{array,multirow}
\usepackage{tabularx}
\usepackage{courier}
\usepackage{url}
\usepackage{booktabs}
\usepackage[final]{pdfpages}
\usepackage{bbold}
\usepackage{tikz}
\usepackage{mathtools}
\usepackage{bm}
\usepackage{mathtools}

\usepackage{amssymb}
\usepackage{amsmath}
\usepackage[usestackEOL]{stackengine}
\usepackage{mathtools}
\usepackage{afterpage}
\usepackage{seqsplit}
\usepackage[page,toc,titletoc,title]{appendix}
\usepackage[export]{adjustbox}
\usepackage{authblk}
\usepackage{setspace}
\usepackage[top=1.2in, bottom=1.2in]{geometry}

\makeatletter
\newsavebox{\@brx}
\newcommand{\llangle}[1][]{\savebox{\@brx}{\(\m@th{#1\langle}\)}%
  \mathopen{\copy\@brx\kern-0.5\wd\@brx\usebox{\@brx}}}
\newcommand{\rrangle}[1][]{\savebox{\@brx}{\(\m@th{#1\rangle}\)}%
  \mathclose{\copy\@brx\kern-0.5\wd\@brx\usebox{\@brx}}}
\makeatother

\DeclareMathAlphabet{\mathpzc}{OT1}{pzc}{m}{it}

\newtheorem{innercustomgeneric}{\customgenericname}
\providecommand{\customgenericname}{}
\newcommand{\newcustomtheorem}[2]{%
  \newenvironment{#1}[1]
  {%
   \renewcommand\customgenericname{#2}%
   \renewcommand\theinnercustomgeneric{##1}%
   \innercustomgeneric
  }
  {\endinnercustomgeneric}
}
\newcustomtheorem{lem}{Lemma}
\newcustomtheorem{pos}{Postulate}
\newcustomtheorem{thm}{Theorem}

\newtheorem{theorem}{Theorem}
\newtheorem{lemma}{Lemma}

{
  \begin{center}
  \par\vspace{\baselineskip}\noindent
}%
{
\end{center}
\par\addvspace{\baselineskip}
}

{
  \begin{center}
}%
{
\end{center}
\par\addvspace{\baselineskip}
}

\definecolor{codegreen}{rgb}{0,0.6,0}
\definecolor{cream}{rgb}{1.0,0.99,0.82}
\definecolor{codegray}{rgb}{0.5,0.5,0.5}
\definecolor{cornsilk}{rgb}{1.0, 0.97, 0.86}
\definecolor{codepurple}{rgb}{0.58,0,0.82}
\definecolor{backcolour}{rgb}{0.95,0.95,0.92}
\definecolor{athena}{rgb}{0.40, 0.20, 0.70}
\lstdefinestyle{mystyle}{
    commentstyle=\color{codegreen},
    numberstyle=\tiny\color{codegray},
    stringstyle=\color{codepurple},
    basicstyle=\scriptsize\ttfamily,
    breakatwhitespace=false,         
    breaklines=true,
    keywordstyle=\bfseries\color{Blue},
    alsoletter=!-,  
    captionpos=b,                    
    keepspaces=true,     
	mathescape=true,
    frameround=tttt,
    rulecolor=\color{Bittersweet}, 
    numbersep=1pt,                  
    showstringspaces=false,
    showtabs=false,                  
    tabsize=2 
} 
\renewcommand{\top}{\intercal}

\lstdefinestyle{athena-list}{
    commentstyle=\color{codegreen},
    numberstyle=\tiny\color{codegray},
    stringstyle=\color{codepurple},
    basicstyle=\scriptsize\ttfamily,
    breakatwhitespace=false,         
    breaklines=true,
    keywordstyle=\bfseries\color{athena},
    alsoletter=!-,  
    morekeywords={load,datatype,domain,declare,define,conclude,pick-any,assume,let},                 
    captionpos=b,                    
    keepspaces=true,     
	mathescape=true,
    frame = single,
    frameround=tttt,
    rulecolor=\color{red}, 
    numbersep=1pt,                  
    showstringspaces=false,
    showtabs=false,                  
    tabsize=2,
    escapechar=\@
}

\makeatletter
\newcommand\aepath[1]{
  \bgroup
    \ttfamily
    \ae@path#1\relax\@nil
  \egroup}
\def\ae@path#1#2\@nil{
  \def\ae@continue{}
  \detokenize{#1}\unskip\penalty\z@  
  \ifx\relax#2
  \else 
    \def\ae@continue{\ae@path#2\@nil}
  \fi
  \ae@continue}
\makeatother

\let\mytexttt\aepath

\newcommand{\config}[3]{\llangle #1 \; \; \| \; \; #2 \; \; \| \; \; #3 \rrangle}

\newcommand{\msg}[2]{\langle #1 \Leftarrow #2 \rangle}
\newcommand{\lambdat}{\rightarrow_\lambda}
\newcommand{\trule}[2]{\frac{{\displaystyle #1}}{{\displaystyle #2}}}

\newcommand{\send}{\mbox{{\tt send}}}
\newcommand{\new}{\mbox{{\tt new}}}
\newcommand{\ready}{\mbox{{\tt ready}}}

\newcommand{\funl}{\mathrm{\mathbf{fun}}}
\newcommand{\newl}{\mathrm{\mathbf{new}}}
\newcommand{\sendl}{\mathrm{\mathbf{snd}}}
\newcommand{\receivel}{\mathrm{\mathbf{rcv}}}
\newcommand{\stopl}{\mathrm{\mathbf{stp}}}
\newcommand{\beginl}{\mathrm{\mathbf{bgn}}}

\newcommand{\mapupdate}[3]{#1,[{#3}]_{#2}}  
\newcommand{\alabelt}[1]{ 
  \mathrel{\stackrel{#1}{\longrightarrow}}  }

\newcommand{\multisetunion}[2]{#1 \uplus #2}
\newcommand{\multisetupdate}[2]{\multisetunion{#1}{\{ #2 \}}}

\newcommand{\redcontext}{\mathsf{R}}
\newcommand{\redex}{\mathsf{r}}
\newcommand{\restrict}[2]{#1|_{\mbox{\fontfamily{lmss}\selectfont{\em {dom}}}(#1)-\{#2\}}}
\newcommand{\freevar}[1]{\mbox{\fontfamily{lmss}\selectfont{\em {fv}}}({#1})}
\newcommand{\domain}[1]{\mbox{\fontfamily{lmss}\selectfont{\em {dom}}}({#1})}

\newcommand{\conred}[2]{#1\blacktriangleright#2\blacktriangleleft}
\newcommand{\redexp}[1]{\conred{\redcontext}{#1}}

\newcommand{\nil}{\mbox{{\tt nil}}}

\begin{document}





\begin{titlepage}
\begin{center}

\begin{spacing}{1.5}
\bigskip
\bigskip
\bigskip
\bigskip
\textbf{\Large Verification of Eventual Consensus in Synod Using a Failure-Aware Actor Model}\\[0.5cm]
{Saswata Paul\textsuperscript{1}, Gul A. Agha\textsuperscript{2}, Stacy Patterson\textsuperscript{1}, and Carlos A. Varela\textsuperscript{1}}\\
{\small \textsuperscript{1}Rensselaer Polytechnic Institute, Troy, New York, 12180, USA}\\
{\small {\texttt{pauls4@rpi.edu, \{sep,cvarela\}@cs.rpi.edu}}}\\
{\small \textsuperscript{2}University of Illinois at Urbana-Champaign, Champaign, Illinois, 61820, USA}
{\small {\texttt{agha@illinois.edu}}}\\
\end{spacing}
\bigskip
\bigskip
\bigskip
\bigskip
\bigskip
\bigskip
\bigskip
\bigskip
\bigskip
\bigskip
\bigskip
\bigskip
\bigskip
\begin{spacing}{}

\textbf{\Large Technical Report\textsuperscript{*}\\}
\today \\

\vspace*{\fill}

\end{spacing}

{\let\thefootnote\relax\footnote{{\textsuperscript{*}This technical report is an extended version of the \textbf{13\textsuperscript{th} NASA Formal Methods Symposium (May 24-28, 2021)} proceedings paper with the same name. It contains the Athena codes for the formal proofs presented in the proceedings paper.}}}
\end{center}
\end{titlepage}

\input{abstract}

\tableofcontents

\newpage
\input{intro}

\input{synod}

\input{actor}

\input{proof}

\input{related}

\input{conc}

\input{acknowledge}

\bibliographystyle{splncs04}
\bibliography{references}

\input{appendices}

\end{document}

%% file: abstract.tex
\begin{abstract}

Successfully attaining consensus in the absence of a centralized coordinator is a fundamental problem in distributed multi-agent systems.
We analyze progress in the Synod consensus protocol---which does not assume a unique leader---under the assumptions of asynchronous communication and potential agent failures.
We identify a set of sufficient conditions under which it is possible to guarantee that a set of agents will eventually attain consensus.
First, a subset of the agents must behave correctly and not permanently fail until consensus is reached, and second, at least one proposal must be eventually uninterrupted by higher-numbered proposals.  
To formally reason about agent failures, we introduce a failure-aware actor model (FAM).
Using FAM, we model the identified conditions and provide a formal proof of eventual progress in Synod.
Our proof has been mechanically verified using the Athena proof assistant and, to the best of our knowledge, it is the first machine-checked proof of eventual progress in Synod.


\end{abstract}

%% file: intro.tex
\section{Introduction}\label{intro}

\emph{Consensus}, which requires a set of processes to reach an agreement on some value, is a fundamental problem in distributed systems.
Under \emph{asynchronous} communication settings, where message transmission and processing delays are unbounded, it is impossible to guarantee consensus~\cite{fischer1985impossibility} since message delays cannot be differentiated from process failures.
Nevertheless, in distributed \emph{multi-agent} systems, where there is no centralized \emph{coordinator} to manage safe operation, it is necessary for the agents to use \emph{distributed consensus protocols}~\cite{ren2008distributed} for coordination.
An important application of such systems is decentralized \emph{air-traffic control} (ATC) for \emph{Urban Air Mobility} (UAM)~\cite{vascik2018analysis}.

The integration of \textit{uncrewed aircraft systems} (UAS) and \emph{micro-aircraft} in the \textit{National Airspace System} (NAS) for package delivery, scientific data collection, and urban transportation will significantly increase the density of urban air traffic~\cite{NAP25143}, elevating the possibilities of hazards such as \emph{near mid-air collisions} (NMAC)~\cite{lee2013investigating} and \emph{wake-vortex induced rolls}~\cite{luckner2004hazard}.
Since centralized, human-operated ATC is not scalable to high densities and is prone to human errors~\cite{hopkin2017human}, UAS operating in UAM scenarios must be capable of autonomous \emph{UAS traffic management} (UTM)~\cite{aweiss2018unmanned}.
To ensure safety in UTM, they must coordinate with each other by using distributed consensus protocols\footnote{An approach for \emph{implicit coordination} between two aircraft has been proposed in \cite{narkawicz2016coordination}, but that is only applicable for the purpose of \emph{pairwise tactical conflict avoidance}.} (\emph{e.g.} --~\cite{balachandran2018distributed, peters2020flight}).

In~\cite{paul-dddas-2020}, we have proposed an \emph{Internet of Planes} (IoP), consisting of an asynchronous \emph{vehicle-to-vehicle} (V2V)~\cite{molisch2009survey} network of aircraft, to facilitate autonomous capabilities such as \emph{decentralized admission control} (DAC)~\cite{paul-dasc-2020}.
In DAC, a \emph{candidate} aircraft generates
a \emph{conflict-aware flight plan} that avoids NMACs with a set of \emph{owner} aircraft of a \emph{controlled airspace}~\cite{paul-dasc-2019}.
The candidate then requests admission into the airspace by proposing this flight plan to the owners.  
As there may be multiple candidates concurrently competing for admission into an airspace, the owners may only admit
candidates sequentially.  
This is because candidates do not consider each other in their proposals: in fact, a
candidate may not even be aware of other candidates.  
Thus, admitting one candidate potentially invalidates the proposals of all other
candidates.
Since UAM applications are \emph{time-critical}, any consensus protocol used for DAC must guarantee that a proposal will be eventually chosen.
In this paper, we present an analysis of consensus that is primarily motivated by the requirements of DAC.

In \cite{lamport1998part}, Lamport describes three
variants of a consensus protocol that strongly guarantees the
\emph{safety} property that only one value will be chosen:
\begin{itemize}
\item \emph{The Basic Synod protocol} or \emph{Synod} guarantees safety if one or more agents are allowed to initiate new proposals for consensus.

\item \emph{The Complete Synod protocol} or \emph{Paxos} is a variant of Synod in which only a distinguished agent (\emph{leader}) is allowed to initiate proposals~\cite{lamport2001paxos}. 
Other agents may only introduce values through the leader.
This is done to guarantee the \emph{progress} property that consensus will eventually be achieved. 

\item \emph{The Multi-Decree protocol} or \emph{Multi-Paxos} allows multiple values to be chosen using a separate instance of Paxos for each value, but using the same leader for each of those instances.
\end{itemize}
Both Paxos and Multi-Paxos use Synod as the underlying consensus protocol.

A progress guarantee contingent upon a unique leader has several drawbacks from the perspective of DAC. 
First, \emph{leader election} itself is a consensus problem.
Therefore, a progress guarantee that relies on successful leader election as a precondition would be circular, and therefore fallacious.
Second, the Fischer, Lynch, and Paterson \emph{impossibility result} (FLP)~\cite{fischer1985impossibility} implies that unique leadership cannot be guaranteed in asynchronous systems where agents may unpredictably fail.
In some cases, network partitioning may also erroneously cause multiple leaders to be elected~\cite{alquraan2018analysis}.
Third, V2V networks like the IoP are expected to be highly dynamic where membership frequently changes.
Consensus is used in DAC for agreeing on only a single candidate, after which the set of owners changes by design.
Hence, the benefits of electing a stable leader that apply for \emph{state machine replication} (\emph{e.g.} -- \cite{kirsch2008paxos}), where reconfiguration is expected to be infrequent~\cite{lamport2009vertical}, are not applicable to DAC.
Finally, channeling proposals through a unique leader creates a communication bottleneck and introduces the leader as a single point of failure: there can be no progress if the specified leader fails.  
In the absence of a unique leader, a system implementing Paxos falls back to the more general Synod protocol.  
Therefore, in this paper, we focus on identifying sufficient conditions under which the fundamental Synod protocol can make eventual progress.

The failure of \emph{safety-critical} aerospace systems can lead to the loss of human life~\cite{imai-varela-pilots-dddas-2012, sommerville2011software} and property~\cite{imai-pilots-ieee-aesm-2017}.  
Hence, formal methods must be used for the rigorous verification of any algorithm used in such systems.
The \emph{actor model}~\cite{agha1986actors, hewitt1977viewing} is a theoretical model of concurrent computation that can be used for formal reasoning about distributed algorithms~\cite{musser-varela-agere-2013}.
It assumes asynchronous communication as the most primitive form of interaction and it also assumes \emph{fairness}, which is useful for reasoning about the progress of actor systems~\cite{varela2013programming}.
In the context of DAC, aircraft may experience temporary or permanent communication failures where they are unable to send or receive any messages~\cite{okcu2016operational}. 
Our verification of Synod must model the possibility of such failures.  
To support explicit reasoning about such failures in actors, we introduce a \emph{(predicate-fair) failure-aware actor model} (FAM) that assumes \textit{predicate fairness}~\cite{queille1983fairness} in addition to the standard actor model's fairness properties. 
Predicate fairness states that if a predicate is \emph{enabled infinitely often} in a given path, then it must be eventually satisfied.   
We use Varela's dialect~\cite{varela2013programming}
of Agha, Mason, Smith, and Talcott's actor language (AMST)~\cite{agha1997foundation} and modify its semantics to model failures.

In order to ensure that our formalization of Synod and its eventual progress property are correct, we machine-check our proof using the \emph{Athena proof assistant}~\cite{athena1, arkoudas2017fundamental}.
Along with a language for expressing proofs, Athena also provides an interactive proof development environment.
Athena's theorem proving capabilities are based on \emph{many-sorted first-order logic}~\cite{manzano1996extensions} and it uses a \emph{natural deduction}~\cite{arkoudas2005simplifying} style of proofs.
All terms have an associated \emph{sort} and Athena can automatically detect and report \emph{ill-sorted} terms and expressions in proofs.
Athena is \emph{sound}: all methods that successfully execute produce a theorem that is guaranteed to be a logical consequence of its \emph{assumption base}. 
It also allows the use of \emph{automated theorem provers} like Vampire~\cite{riazanov2002design} and SPASS~\cite{weidenbach2009spass}.  

The main contributions of this work are:
\begin{itemize}
\item We identify a set of conditions under which progress in Synod can be guaranteed in purely asynchronous settings, without assuming a unique leader.

\item We introduce a failure-aware actor model to support formal reasoning about temporary or permanent actor failures.

\item We show that progress can be guaranteed in Synod under the identified conditions and mechanically verify our proof using Athena.  
To our knowledge, this is the first machine-checked proof of eventual progress in Synod.

\end{itemize}

It is important to note here that a guarantee of eventual progress \emph{alone} is insufficient for time-critical UAM applications, but it is a necessary precondition for providing \emph{timely progress} guarantees that can be directly applicable for UAM. 


The paper is structured as follows -- 
Section~\ref{synod} informally describes Synod and discusses the conditions required for progress; 
Section~\ref{actor} introduces FAM; 
Section~\ref{proof} presents the formal verification of progress in Synod;
Section~\ref{related} relates our work to prior research on formal verification of consensus protocols;
and Section~\ref{conc} concludes the paper including potential future directions of work.

%% file: synod.tex
\section{The Synod Protocol}\label{synod}

Synod assumes an asynchronous, non-Byzantine system model in which
agents operate at arbitrary speed, may fail and restart, and have
stable storage.  Messages can be duplicated, lost, and have arbitrary
transmission time, but cannot be corrupted\cite{lamport2001paxos}.  It
consists of two logically separate sets of agents:

\begin{itemize}
\item \emph{Proposers} - The set of agents that can propose values to be chosen.
\item \emph{Acceptors} - The set of agents that can vote on which value should be chosen.
\end{itemize}

Synod requires a subset of acceptors, which satisfy a
\textit{quorum}, to proceed.  
To ensure that if there is a consensus in one quorum, there cannot also be another quorum with a consensus, any two quorums must intersect.
A subset of acceptors which constitutes a simple majority is an example of a quorum.
Other methods of determining quorums can be found in~\cite{howard2016flexible}. 

There are four types of messages in Synod: 
\begin{itemize}
\item \emph{prepare (1a)} messages include a \emph{proposal number}. 
\item \emph{accept (2a)} messages include a proposal number and a \emph{value}. 
\item \emph{promise (1b)} messages include a proposal number and a value. 
\item \emph{voted (2b)} messages include a proposal number and a value. 
\end{itemize}

For each proposer, the algorithm proceeds in two distinct
\emph{phases}\cite{lamport2001paxos}:

\begin{itemize}
\item \textbf{Phase 1}
\begin{enumerate}[label={(\alph*)}]

\item A proposer $P$ selects a \emph{unique proposal number} $b$ and sends a \textit{prepare} request with $b$ to a subset $Q$ of acceptors, where $Q$ constitutes a quorum.
\item 
When an acceptor $A$ receives a
  \textit{prepare} request with the proposal number $b$, it checks if $b$ is greater than the proposal numbers of all \textit{prepare} requests to which $A$ has already responded.  If this condition is satisfied, then $A$ responds to $P$ with a \textit{promise} message.  The \emph{promise} message implies that $A$ will not accept any other
  proposal with a proposal number less than $b$. The promise includes (i) the highest-numbered proposal $b'$ that $A$ has previously accepted and (ii) the value corresponding to $b'$.  If~$A$ has not accepted any proposals, it simply sends a default value.
\end{enumerate}

\item \textbf{Phase 2}
\begin{enumerate}[label={(\alph*)}]
\item If $P$ receives a \emph{promise} message in response to its \textit{prepare} requests from all members of~$Q$, then~$P$ sends an \textit{accept} request to all members of~$Q$.  These \emph{accept} messages contain the proposal number~$b$ and a value~$v$, where~$v$ is the value of the highest-numbered proposal among the responses or an arbitrary value if all responses reported the default value. 

\item If an acceptor $A$ receives an \textit{accept} request with a proposal number $b$ and a value $v$ from $P$, it accepts the proposal unless it has already responded to a \textit{prepare} request having a number greater than $b$.  If $A$ accepts the proposal, it sends $P$ a \textit{voted} message which includes $b$ and $v$\cite{lamport1998part}.
\end{enumerate}
\end{itemize}
A proposer \emph{determines} that a proposal was successfully \emph{chosen} if and only if it receives \textit{voted} messages from a quorum for that proposal.

Synod allows multiple proposals to be chosen. 
Safety is ensured by the invariant - \emph{"If a proposal with value $v$ is chosen, then every higher numbered-proposal that is chosen has value $v$"}\cite{lamport2001paxos}. 
Therefore, there may be situations where a proposer $P$ proposes a proposal number after one or more lower-numbered proposals have already been chosen, resulting in some value $v$ being chosen. 
By design, Synod will ensure that $P$ proposes the same value $v$ in its Phase 2.  
Since proposers can initiate proposals in any order and communication is asynchronous, the only fact that can be guaranteed about any chosen value is that it must have been proposed by the first proposal to have been chosen by any quorum.   
From the context of admission control, it implies that if a candidate $P_2$ successfully completes both phases after another candidate $P_1$ has completed both phases, then $P_2$ will simply learn that $P_1$ has been granted admission. 
So $P_2$ will update its set of owners to include $P_1$, create a new conflict-aware flight plan, and request admission by starting the Synod protocol again.

\subsection{Progress in Synod}\label{progress}

Some obvious scenarios in which progress may be affected in Synod are:
\begin{itemize}
\item  
Two proposers $P_1$ and $P_2$ may complete Phase 1 with proposal numbers $b_1$ and $b_2$ such that $b_2> b_1$. 
This will cause  $P_1$ to fail Phase 2.
$P_1$ may then propose a fresh proposal number $b_3 > b_2$ and complete Phase 1 before $P_2$ completes its Phase 2. 
This will cause $P_2$ to fail Phase 2 and propose a fresh proposal number $b_4 > b_3$ . 
This process may repeat infinitely\cite{lamport2001paxos} (\emph{livelock}).

\item 
Progress may be affected even if one random agent fails unpredictably~\cite{fischer1985impossibility}. 
  
\end{itemize}

Paxos assumes that a distinguished proposer (leader) is elected as the only proposer that can initiate proposals\cite{lamport1998part}.
Lamport states \emph{``If the distinguished proposer can communicate successfully with a majority of acceptors, and if it uses a proposal with number greater than any already used, then it will succeed in issuing a proposal that is accepted.''}\cite{lamport2001paxos}.
The FLP impossibility result\cite{fischer1985impossibility} implies that in purely asynchronous systems, where agent failures cannot be differentiated from message delays, leader election cannot be guaranteed.
Moreover, it is possible that due to network partitioning, multiple proposers are elected as leaders\cite{alquraan2018analysis}.
In the absence of a unique leader, a system implementing Paxos falls back to Synod. 
Therefore, it is important to identify the conditions under which progress can be formally guaranteed in Synod in the absence of a unique leader.

To guarantee progress, it suffices to show that \emph{some} proposal number $b$ will be chosen.
This will happen if $b$ satisfies the following conditions: 
\begin{enumerate}[label=\textit{{P\arabic*}}]
\item When an acceptor $A$ receives a \emph{prepare} message with $b$, $b$ should be greater than all other proposal numbers that $A$ has previously seen.
\item When an acceptor $A$ receives an \emph{accept} message with $b$, $b$ should be greater than or equal to all other proposal numbers that $A$ has previously seen. 
\end{enumerate}
\textit{P1} and \textit{P2} simply suggest that for $b$, there will be a long enough period without \emph{prepare} or \emph{accept} messages with a proposal number greater than $b$, allowing messages corresponding to $b$ to get successfully processed without being interrupted by messages corresponding to a higher-numbered proposal.  
The non-interruption condition follows from two assumptions.  
First, we assume that a proposer $P$ will keep retrying until it successfully receives votes from a quorum. 
Second, we assume that the system has a form of fairness called \emph{predicate fairness}~\cite{queille1983fairness}. 
With predicate fairness, if a predicate is enabled infinitely often in a given path, then the predicate must be satisfied (this is a recursive definition in that the path could begin from any state along the path). 
Conditions \textit{P1} and \textit{P2} are infinitely often enabled in any path corresponding to livelock.
Therefore, by predicate fairness, these conditions must eventually happen, allowing consensus to be reached.\footnote{We do not use the predicate fairness assumption in the formal proof associated with this paper. Instead, we use a system-specific derived property: that eventually, at least one proposal must be uninterrupted by higher-numbered proposals.}

Since progress cannot be guaranteed if too many agents permanently fail or if too many messages are lost, Lamport~\cite{lamport2006fast} presents some conditions for informally proving progress in Paxos.
A \emph{nonfaulty} agent is defined as \emph{``an agent that eventually performs the actions that it should''}, and a \emph{good} set is defined as a set of nonfaulty agents, such that, if an agent repeatedly sends a message to another agent in the set, it is eventually received by the recipient.
It is then assumed that the unique leader and a quorum of acceptors form a good set and that they infinitely repeat all messages that they have sent. 
These conditions are quite strong since they depend on the future behavior of a subset of agents and may not always be true of an implementation.
However, since they have been deemed reasonable for informally proving progress even in the presence of a unique leader, we partially incorporate them in our conditions under which progress in Synod can be formally guaranteed in the absence of a unique leader.
Our complete set of conditions for guaranteeing progress in Synod, therefore, informally states that \emph{eventually, a nonfaulty proposer must propose a proposal number, that will satisfy P1 and P2, to a quorum of nonfaulty acceptors, and the Synod-specific messages between these agents must be eventually received}.

We can see that the conditions for progress in Paxos constitute a special case of our conditions for progress in Synod where \emph{P1} and \emph{P2} are satisfied by a proposal proposed by the unique leader. 
If the unique leader permanently fails, then the corresponding guarantee is only useful if leader re-election is successful.
However, if leader election is assumed to have already succeeded, there will be no need for further consensus, rendering the guarantee moot.
Synod's progress guarantee remains useful as long as at least one random proposer is available to possibly propose at least one successful proposal, thereby remaining pertinent even if multiple (not all) proposers arbitrarily fail.
Moreover, the conditions do not assume that consensus (leader election) will have already succeeded.

%% file: actor.tex
\section{A Failure-Aware Actor Model}\label{actor}

We use the actor model to formally reason about progress in Synod since it assumes asynchronous communication and fairness, which is helpful for reasoning about progress\cite{agha1997foundation}. 
Fairness in the standard actor model has the following consequences\cite{varela2013programming}:
\begin{itemize}
\item guaranteed message delivery\footnote{"Delivery" here implies that the message will be available to the recipient. The recipient may or may not eventually receive and process the message.}, and 
\item an actor infinitely often ready to process a message will eventually process the message.
\end{itemize}

The IoP is an open network in which aircraft communicate asynchronously and may experience permanent or temporary communication failures that render them unable to send or receive any messages. 
All messages to and from an aircraft may get delayed because of transmission problems or internal processing delays (or processing failures) in the aircraft.
In asynchronous communication, since message transmission times are unbounded, it is not possible to distinguish transmission delays from processing delays or failures.  
However, it is important to take into account if an actor has failed at any given time, \emph{i.e.}, if it is incapable of sending or receiving messages.
For this reason, we introduce a (predicate-fair) failure-aware actor model (FAM) that allows reasoning about such actor failures.

FAM models two states for an actor at any given time---\emph{available} or \emph{failed}.
Actors can switch states as \emph{transitions} between \emph{configurations}.
From the perspective of message transmission and reception, a failed actor cannot send or receive \emph{any} messages, but an available actor can.
The failure model of FAM also assumes that every actor has a stable storage that is persistent across failures.
In addition to the standard actor model's fairness assumptions, FAM assumes predicate fairness\cite{queille1983fairness}, which states that a predicate that is infinitely often enabled in a given path will eventually be satisfied.

Varela~\cite{varela2013programming} presents a dialect of AMST's \emph{lambda-calculus} based actor language~\cite{agha1997foundation} whose operational semantics are a set of \emph{labeled transitions} from \emph{actor configurations} to actor configurations\footnote{Interested readers can refer to section 4.5 of \cite{varela2013programming} for more details.}.
An actor configuration $\kappa$ is a temporal snapshot of actor system components, namely the individual actors and the messages ``en route''.
It is denoted by $\llangle \alpha \ \ \| \ \ \mu \rrangle$,
where $\alpha$ is a map from actor names to \emph{actor expressions}, and $\mu$ is a multi-set of messages. 
An actor expression $e$ is either a value $v$ or a \emph{reduction context} $\redcontext$ filled with a \emph{redex} $\redex$, denoted as $e = \conred{\redcontext}{\redex}$.
$\kappa_1 \xrightarrow{l} \kappa_2 $ denotes a \emph{transition rule} where $\kappa_1$, $\kappa_2$, and $l$ are the initial configuration, the final configuration, and the transition label respectively. 
There are four possible transitions -- $\funl$, $\newl$, $\sendl$, and $\receivel$. 
To model failures, we modify Varela's dialect of AMST and categorize its original transitions ($\funl$, $\newl$, $\sendl$, and $\receivel$) as \emph{base-level} transitions. 

For a base-level transition in FAM to be enabled to occur for an \emph{actor in focus} at any time, the actor needs to be available at that time.
To denote available and failed actors at a given time, we redefine an actor configuration as $\llangle \alpha \ \ \| \ \ \bar{\alpha}   \ \ \| \ \ \mu \rrangle$,
where $\alpha$ is a map from actor names to actor expressions for available actors, $\bar{\alpha}$ is a map from actor names to actor expressions for failed actors, and $\mu$ is a multi-set of messages ``en route''.

To model actor failure and restart, we define two \emph{meta-level} transitions $\stopl$ (\emph{stop}) and $\beginl$ (\emph{begin}) that can stop an available actor or start a failed actor in its persistent state before failure.
The $\stopl$ transition is only enabled for an actor in the available state and the $\beginl$ transition is only enabled for an actor in the failed state.
Fig.~\ref{fig:base} and Fig.~\ref{fig:meta} show the operational semantics of our actor language as labelled transition rules\footnote{
$\rightarrow_{\lambda}$ denotes lambda calculus semantics, essentially beta-reduction.
\texttt{new}, \texttt{send}, and \texttt{ready} are actor redexes. 
$\msg{a}{v}$ denotes a message for actor $a$ with value $v$. $\mapupdate{\alpha}{a}{e}$ denotes the extended map $\alpha'$, which is the same as $\alpha$ except that it maps $a$ to $e$. $\multisetunion{}{}$~denotes multiset union.
$\alpha|_S$ denotes restriction of mapping $\alpha$ to elements in set $S$.
$\domain{\alpha}$ is the domain of $\alpha$.
}$^{,}$\footnote{More details about actor language semantics can be found in  \cite{agha1992towards}, \cite{agha1997foundation}, and \cite{varela2013programming}.}.

\begin{figure}[t]
\centering
\begin{equation*}
\trule{e \lambdat e'}
                  {\config{\mapupdate{\alpha}{a}{\redexp{e}}}{\bar{\alpha}}{\mu}
                    \alabelt{[\funl: a]}
                   \config{\mapupdate{\alpha}{a}{\redexp{e'}}}{\bar{\alpha}}{\mu}}
\end{equation*}   
\begin{equation*}
\config{\mapupdate{\alpha}{a}{\redexp{\new(b)}}}{\bar{\alpha}}{\mu}
                    \alabelt{[\newl: a,a']}
                   \config{\mapupdate{\mapupdate{\alpha}{a}{\redexp{a'}}}{a'}{\ready(b)}}{\bar{\alpha}}{\underset{\mbox{{\em $a'$ fresh}}}{\mu}}   
\end{equation*}   
\begin{equation*}
\config{\mapupdate{\alpha}{a}{\redexp{\send(a',v)}}}
              {\bar{\alpha}}{\mu}
             \alabelt{[\sendl: a]}
             \config{\mapupdate{\alpha}{a}{\redexp{\nil}}}
              {\bar{\alpha}}{\multisetupdate{\mu}{\msg{a'}{v}}}
\end{equation*} 
\begin{equation*}
\config{\mapupdate{\alpha}{a}{\redexp{\ready(b)}}}{\bar{\alpha}}
              {\multisetunion{\{\msg{a}{v}\}}{\mu}}
             \alabelt{[\receivel: a,v]}
             \config{\mapupdate{\alpha}{a}{b(v)}}{\bar{\alpha}}{\mu}  
\end{equation*}  
\captionsetup{belowskip=0pt}
\caption{Operational semantics for the base-level transition rules.}
\label{fig:base}

\centering
\begin{equation*}
\config{\mapupdate{\alpha}{a}{e}}
              {\bar{\alpha}}{\mu}
             \alabelt{[\stopl: a]}
             \config{\restrict{\alpha}{a}}
              {\mapupdate{\bar{\alpha}}{a}{e}}{\mu}
\end{equation*}
\begin{equation*}
\config{\alpha}
              {\mapupdate{\bar{\alpha}}{a}{e}}{\mu}
             \alabelt{[\beginl: a]}
             \config{\mapupdate{\alpha}{a}{e}}
              {\restrict{\bar{\alpha}}{a}}{\mu}
\end{equation*}

\captionsetup{belowskip=0pt}
\caption{Operational semantics for the meta-level transition rules.}
\label{fig:meta}
\end{figure}

For an actor configuration $\kappa = \llangle \alpha \ \ \| \ \ \bar{\alpha}   \ \ \| \ \ \mu \rrangle$ to be syntactically well-formed in our actor language, it must conform to the following\footnote{$\freevar{e}$ is the set of free variables in the expression $e$.}:
\begin{enumerate}
    \item $\forall a$, $a \in \domain{\alpha} \cup \domain{\bar{\alpha}}$, $\freevar{\alpha(a)} \subseteq  \domain{\alpha} \cup \domain{\bar{\alpha}}$ 
    \item $\forall m$, $m \in \mu$, $m = \msg{a}{v}$, $\freevar{a} \cup \freevar{v} \subseteq \domain{\alpha} \cup \domain{\bar{\alpha}}$
    \item $\domain{\alpha} \cap \domain{\bar{\alpha}} = \emptyset$ 
\end{enumerate}

The standard actor model's fairness assumptions apply only to the base-level transitions in our language and not to the meta-level transitions.

%% file: proof.tex
\section{Formal Verification of Eventual Progress in Synod}\label{proof}

This section presents our proof of eventual progress in the Synod protocol. 
The notations used in this section have been introduced in Table~1 and Table~2.

\begin{table}[t]
\centering

\begin{adjustbox}{width=\textwidth}

\begin{tabular}{|c|c|c|c|}
\hline
\textbf{Symbol} & \textbf{Description} & \textbf{Symbol} & \textbf{Description} \\ \hline
$\mathcal{A}$       & Set of all actors  & $\mathcal{M}$       & Set of all messages            \\ \hline                               
$\mathbb{P}$       & Set of all proposer actors  & $\mathbb{A}$       & Set of all acceptor actors            \\ \hline                               
$\mathcal{V}$       & Set of all values  & $\mathcal{Q}$       & Set of all quorums            \\ \hline                     
$\mathcal{B}$       & Set of all proposal numbers  & $\mathbb{M}$       & Set of all sets of messages                      
\\ \hline
       
$\mathcal{C}$       & Set of all actor configurations  & $\mathcal{S}$       & Set of all transition steps            \\ \hline                                          
$\mathcal{T}$       & Set of all fair transition paths    &  $\mathbb{N}$ & Set of all natural numbers        \\ \hline

\end{tabular}

\end{adjustbox}

\label{tab:doms}
\caption{Set symbols for our formal specification.}

\centering

\begin{adjustbox}{width=\textwidth}

\begin{tabular}{|c|c|c|c|}
\hline
\textbf{Symbol} & \textbf{Description} & \textbf{Input} & \textbf{Output} \\ \hline
  $\varsigma$     &  Get last configuration           &  $\mathcal{T}$     &  $\mathcal{C}$      \\ \hline
   $\rho$    &    Get transition path up to index         &  $\mathcal{T} \times \mathbb{N}$     &   $\mathcal{T}$     \\ \hline
$\top$      &  Transition path constructor         &  $\mathcal{T}  \times \mathcal{S}$     & $\mathcal{T}$       \\ \hline
$\sigma$      &  Choose a value to propose based on configuration     &  $\mathcal{C} \times \mathcal{A}$     & $\mathcal{V}$       \\ \hline
$\mathfrak{s}$      &  Construct a \texttt{snd} transition step     &  $\mathcal{A} \times \mathcal{M}$     & $\mathcal{S}$       \\ \hline
$\mathfrak{r}$      &  Construct a \texttt{rcv} transition step     &  $\mathcal{A} \times \mathcal{M}$     & $\mathcal{S}$       \\ \hline
$\mathfrak{m}$      &  Get set of messages ``en route''      &  $\mathcal{C}$     & $\mathbb{M}$       \\ \hline
$\mathfrak{a}$      &  Actor is in $\alpha$    &  $\mathcal{C} \times \mathcal{A}$     & \texttt{Bool}       \\ \hline
$\Re$      &  Actor is ready for a step           &  $\mathcal{T} \times \mathcal{A} \times \mathcal{S}$     & \texttt{Bool}       \\ \hline
$\phi$      &  Proposer has promises from a quorum         &  $\mathcal{A} \times \mathcal{B} \times \mathcal{Q} \times \mathcal{C} $     & $\texttt{Bool}$       \\ \hline
$\Phi$      &  Proposer has votes from a quorum            &  $\mathcal{A} \times \mathcal{B} \times \mathcal{Q} \times \mathcal{C}  $     & $\texttt{Bool}$       \\ \hline
$\dj$      &  Actor is nonfaulty     &  $\mathcal{A}$     & $\texttt{Bool}$       \\ \hline
$\th$      &  Proposal number satisfies \emph{P1} and \emph{P2}           &  $\mathcal{B}$     & $\texttt{Bool}$       \\ \hline
$\L$      &  Proposer has learned of successful consensus            &  $\mathcal{A} \times \mathcal{B} \times \mathcal{C}  $     & $\texttt{Bool}$       \\ \hline
\end{tabular}

\end{adjustbox}

\label{tab:funs}
\caption{Relation symbols for our formal specification.}
\end{table}

A message is a tuple $\langle \texttt{s} {\in}\mathcal{A} , \texttt{r} {\in} \mathcal{A}, \texttt{k} {\in} \xi, \texttt{b} {\in} \mathcal{B}, \texttt{v} {\in} \mathcal{V}  \rangle$ where $\xi = \{ 1a, 1b, 2a , 2b \} $, \texttt{s}~is the sender, \texttt{r} is the receiver, \texttt{k} is the type of message, \texttt{b} is a proposal number, and \texttt{v} is a value.
$\bar{v} \in \mathcal{V}$ is a null value constant used in $1a$ (\emph{prepare}) messages.

The local state of an actor $x$ can be extracted from a configuration $\kappa$ as a tuple $\langle \eta^x_{\kappa}{\in}\mathbb{M}, \beta^x_{\kappa}{\in}\mathcal{B}, v^x_{\kappa}{\in}\mathcal{V} \rangle$ where $\eta^x_{\kappa}$ is the set of messages received but not yet responded to, $\beta^x_{\kappa}$ is the highest proposal number seen, 
and $v^x_{\kappa}$ is the value corresponding to the highest proposal number accepted.

\emph{Transition paths} represent the dynamic changes to actor configurations as a result of transition steps\cite{musser-varela-agere-2013}.
\emph{Indexed positions} in transition paths correspond to logical steps in time and are used to express eventuality.

$\mathbb{P} \subset \mathcal{A}$ and $\mathbb{A} \subset \mathcal{A}$ are the sets of proposers and acceptors respectively and a quorum is a possibly equal non-empty subset of $\mathbb{A}$, \emph{i.e.},$\forall Q \in \mathcal{Q}: Q \subseteq \mathbb{A} \ \land \ Q \neq \emptyset$.

\subsection{Fairness Assumptions for Transitions}
We assume two fairness axioms for the $\sendl$ and $\receivel$ transitions that follow from the fairness assumptions of FAM.
The \texttt{F-Snd-axm} and the \texttt{F-Rcv-axm} state that if a $\sendl$ or $\receivel$ transition is enabled at some time, it must either eventually happen or eventually, it must become permanently disabled.

\begin{lstlisting}[mathescape=true, basicstyle=\footnotesize\ttfamily]
F-Snd-Axm $\equiv$ 
$ \forall x \in \mathcal{A}, m \in \mathcal{M}, T \in \mathcal{T}, i \in \mathbb{N} : ( \Re(\rho(T,i), x, \mathfrak{s}(x,m)) \implies $ 
								$((\exists j \in \mathbb{N} : (j \geq i)$  
											$ \land \ \rho (T,j+1) = \top (\rho(T,j),\mathfrak{s}(x,m)))$ 
									$\lor \ (\exists k \in \mathbb{N} : (k > i)$  
											$ \land \ (\forall j \in \mathbb{N}: (j \geq k) \implies  \lnot \Re(\rho(T,j), x, \mathfrak{s}(x,m)) ))))$

F-Rcv-Axm $\equiv$ 
$ \forall x \in \mathcal{A}, m \in \mathcal{M},T \in \mathcal{T}, i \in \mathbb{N} : ( (m \in \mathfrak{m}(\varsigma(\rho(T,i))) \ \land \ \Re(\rho(T,i), x, \mathfrak{r}(x,m))) \implies $ 
    		$((\exists j \in \mathbb{N} : (j \geq i)$  
    					$ \land \ \rho (T,j+1) = \top (\rho(T,j),\mathfrak{r}(x,m)))$ 
    			$\lor \ (\exists k \in \mathbb{N} : (k > i)$
                $ \land \ (\forall j \in \mathbb{N} : (j \geq k) \implies$
            				    	$ \lnot ( m \in \mathfrak{m}(\varsigma(\rho(T,j))) \ \land \ \Re(\rho(T,j), x, \mathfrak{r}(x,m)) ) ))))$
\end{lstlisting}

\subsection{Rules Specifying the Actions of Synod Actors}\label{axiom_behavior}
The Synod protocol is presented in \cite{lamport1998part} as a high-level abstraction of the behavior of the agents, while leaving out the implementation details to the discretion of the system developers\cite{ongaro2014search}.
We specify rules over actor local states that dictate if an available Synod actor should become ready to send a message.
Since Synod does not specify \emph{when} a proposer should send a \emph{prepare} message, we leave that behavior unspecified. 
For proving progress, we will assume that eventually, a proposer will be ready to send \emph{prepare} messages to a quorum. 

\begin{lstlisting}[mathescape=true, basicstyle=\footnotesize\ttfamily]
Snd-1b-Rul $\equiv$ 
$ \forall  a:\mathbb{A}, p \in \mathbb{P}, T \in \mathcal{T}, i \in \mathbb{N}, b \in \mathcal{B}:$ 
$(\langle p,a,1a,b,\bar{v} \rangle \in \eta^a_{\varsigma(\rho(T,i))} \ \land \ \beta^a_{\varsigma(\rho(T,i))} < b \ \land \ \mathfrak{a}(\varsigma(\rho(T,i)),a)) \implies $
																$\Re(\rho(T,i),a,\mathfrak{s}(a,\langle a,p,1b,b,v^a_{\varsigma(\rho(T,i))} \rangle))$
\end{lstlisting}

\begin{lstlisting}[mathescape=true, basicstyle=\footnotesize\ttfamily]
Snd-2a-Rul $\equiv$ 
$ \forall p \in \mathbb{P}, T \in \mathcal{T}, i \in \mathbb{N}, b \in \mathcal{B}, Q \in \mathcal{Q} :$ 
$(\phi(p,b,Q,\varsigma(\rho(T,i))) \ \land\ \mathfrak{a}(\varsigma(\rho(T,i)),p) ) \implies $
											$(\forall a \in Q : \Re(\rho(T,i),p,\mathfrak{s}(p,\langle p,a,2a,b,\sigma(\varsigma(\rho(T,i)), p) \rangle)))$
\end{lstlisting}

\begin{lstlisting}[mathescape=true, basicstyle=\footnotesize\ttfamily]
Snd-2b-Rul $\equiv$ 
$ \forall  a:\mathbb{A}, p \in \mathbb{P}, T \in \mathcal{T}, i \in \mathbb{N}, b \in \mathcal{B}, v \in \mathcal{V}:$ 
$(\langle p,a,2a,b,v \rangle \in \eta^a_{\varsigma(\rho(T,i))} \ \land \ \beta^a_{\varsigma(\rho(T,i))} \leq b \ \land \ \mathfrak{a}(\varsigma(\rho(T,i)),a))) \implies $
																			$\Re(\rho(T,i),a,\mathfrak{s}(a,\langle a,p,2b,b,v \rangle))$
\end{lstlisting}

Since the response to a message in Synod is a finite set of actions, if there is a message in the multi-set for a Synod actor and the actor is also available, then a receive transition is enabled.
\begin{lstlisting}[mathescape=true, basicstyle=\footnotesize\ttfamily]
Rcv-Rul $\equiv$ 
$ \forall s,r \in \mathcal{A}, T \in \mathcal{T}, i \in \mathbb{N}, k \in \xi, b \in \mathcal{B}, v \in \mathcal{V} :$ 
$(\langle s,r,k,b,v \rangle \in \mathfrak{m}(\varsigma(\rho(T,i))  \ \land \ \mathfrak{a}(\varsigma(\rho(T,i)),r))) \implies \Re(\rho(T,i),r,\mathfrak{r}(r,\langle s,r,k,b,v \rangle))$
\end{lstlisting}

\subsection{Assumptions About the Future Behavior of Agents}\label{axiom_conditions}

To prove progress in Synod, we borrow some assumptions about the future behavior of nonfaulty agents used by Lamport for informally proving progress in Paxos\cite{lamport2006fast}.
It is worth noting that being nonfaulty does not prohibit an agent from temporarily failing.
It simply means that for every action that needs to be performed by the agent, eventually the agent is available to perform the action and the action happens.
In FAM, a nonfaulty actor can be modelled by asserting that an enabled $\sendl$ or $\receivel$ transition for the actor will eventually happen.
However, given the \texttt{F-Snd-Axm} and \texttt{F-Rcv-Axm} axioms, it suffices to assume that for a nonfaulty actor, if a $\sendl$ or $\receivel$ transition is enabled, then it will either eventually occur or it will be infinitely often enabled.
As FAM does not model message loss, any message in the multi-set will persist until it is received.

We introduce a predicate $\dj$ to specify an actor as nonfaulty, such that:
\begin{itemize}
\item $\dj(x)$ implies that the actor $x$ will be eventually available if there is a message ``en route'' that $x$ needs to receive.
\item $\dj(x)$ implies that the actor $x$ will be eventually available if $x$'s local state dictates that it needs to send a message.
\item $\dj(x)$ implies that if a $\sendl$ or $\receivel$ transition is enabled for the actor $x$, it will either eventually occur or it will be infinitely often enabled. 

\end{itemize}
\begin{lstlisting}[mathescape=true, basicstyle=\footnotesize\ttfamily]
Prp-NF-Axm $\equiv$ 
$ \forall p \in  \mathbb{P} : \dj(p) \implies $
	$(\forall b \in \mathcal{B}, k \in \xi, v \in \mathcal{V}, T \in \mathcal{T}, i \in \mathbb{N}, a \in \mathbb{A}, Q \in \mathcal{Q}:$
						$(\phi(p,b,Q,\varsigma(\rho(T,i)))$
						$\lor\ \langle a,p,k,b,v \rangle \in \mathfrak{m}(\varsigma(\rho(T,i)))) \implies$
															$ \ (\mathfrak{a}(\varsigma(\rho(T,i)),p)$
																$\lor\ (\exists j \in \mathbb{N}: (j > i) \ \land \ \mathfrak{a}(\varsigma(\rho(T,j)),p)))) $

Acc-NF-Axm $\equiv$ 
$ \forall a \in \mathbb{A}: \dj(a) \implies $
	$(\forall p \in  \mathbb{P}, k \in \xi, v \in \mathcal{V}, T \in \mathcal{T}, i \in \mathbb{N}, b \in \mathcal{B} :$
					$( (\langle p,a,1a,b,\bar{v} \rangle \in \eta^a_{\varsigma(\rho(T,i))} \ \land \ (\beta^a_{\varsigma(\rho(T,i))} < b) )$
					$\lor\ (\langle p,a,2a,b,v \rangle \in \eta^a_{\varsigma(\rho(T,i))} \ \land \ (\beta^a_{\varsigma(\rho(T,i))} \leq b) )$
					$\lor\  ( \langle p,a,k,b,v \rangle \in \mathfrak{m}(\varsigma(\rho(T,i))) ) ) \implies$
															$ \ (\mathfrak{a}(\varsigma(\rho(T,i)),a)$
																$\lor\ (\exists j \in \mathbb{N}: (j > i) \ \land \ \mathfrak{a}(\varsigma(\rho(T,j)),a))))$
							
NF-IOE-Axm $\equiv$ 
$ \forall x \in \mathcal{A}: \dj(x) \implies$
	$(\forall T \in \mathcal{T}, i \in \mathbb{N}, m \in \mathcal{M}:$
		$( ( m \in \mathfrak{m}(\varsigma(\rho(T,i)))  \ \land \   \Re(\rho(T,i), x, \mathfrak{r}(x,m)) )  \implies $
		$((\exists j \in \mathbb{N} : (j \geq i)$  
					$ \land \ \rho (T,j+1) = \top (\rho(T,j),\mathfrak{r}(x,m)))$ 
			$\lor\ (\forall k \in \mathbb{N} : (k > i)$
            $ \implies \ (\exists j \in \mathbb{N} : (j \geq k)$
        				    	$ \land \ ( m \in \mathfrak{m}(\varsigma(\rho(T,j))) \ \land \ \Re(\rho(T,j), x, \mathfrak{r}(x,m)) ) ))))$
	$\land \ (   \Re(\rho(T,i), x, \mathfrak{s}(x,m))   \implies $
    			$((\exists j \in \mathbb{N} : (j \geq i)$  
    						$ \land \ \rho (T,j+1) = \top (\rho(T,j),\mathfrak{s}(x,m)))$ 
    				$\lor\ (\forall k \in \mathbb{N} : (k > i)$  
    						$ \implies \ (\exists j \in \mathbb{N}: (j \geq k) \ \land \ \Re(\rho(T,j), x, \mathfrak{s}(x,m)) ))))$
\end{lstlisting}

We then introduce a predicate $\th$ that is true of a proposal number if and only if it satisfies the conditions P1 and P2 described informally in Section~\ref{progress}. 

\begin{lstlisting}[mathescape=true, basicstyle=\footnotesize\ttfamily]
P1-P2-Def $\equiv$ 
$ \forall b \in \mathcal{B}: \th(b) \iff $
	$(\forall p \in \mathbb{P}, T  \in \mathcal{T}, i \in \mathbb{N}, v \in \mathcal{V}, a \in \mathbb{A}:$
									$(     (\langle p,a,1a,b,\bar{v} \rangle \in \eta^a_{\varsigma(\rho(T,i))} \implies \beta^a_{\varsigma(\rho(T,i))} < b)  $
									$\land\   (\langle p,a,2a,b,v \rangle \in \eta^a_{\varsigma(\rho(T,i))} \implies \beta^a_{\varsigma(\rho(T,i))} \leq b)  ) )$
\end{lstlisting}

Finally, the conditions for formally guaranteeing progress state that -- \emph{in all fair transition paths, some nonfaulty proposer $p$ will be eventually ready to propose some proposal number $b$, that will satisfy P1 and P2, to some quorum $Q$ whose members are all nonfaulty}.
\begin{lstlisting}[mathescape=true, basicstyle=\footnotesize\ttfamily]
CND $\equiv$ 
$\forall T \in \mathcal{T}: $
	$(\exists i \in \mathbb{N}, p \in \mathbb{P}, b \in \mathcal{B}, Q \in \mathcal{Q}:$ 
				$(\dj(p) \ \land \ \th(b) \ \land \ (\forall a \in Q: (\dj(a) \ \land \ \Re(\rho(T,i), p, \mathfrak{s}(p, \langle p,a,1a,b,\bar{v} \rangle ))) ) ))$
\end{lstlisting}

\subsection{The Proof of Progress}\label{sketch}

To prove progress in Synod, it suffices to prove that eventually, at least one proposal number will be chosen by some quorum (Section~\ref{synod}).
In our set of conditions \texttt{CND}, we have assumed that some proposer $p$ will eventually propose a proposal number $b$, that will satisfy \emph{P1} and \emph{P2}, to a quorum $Q$.
Our proof strategy is to show that eventually, $p$ will learn that $b$ has been chosen by all members of $Q$.
Theorem~\ref{t1} formally states our main progress guarantee while Lemma~\ref{l1} and Lemma~\ref{l2} state progress in Phase 1 and Phase 2 respectively.

\begin{theorem}
\label{t1}
Given \textup{\texttt{CND}}, in all fair transition paths, eventually some proposer $p$ will learn that some proposal number $b$ has been chosen.
\end{theorem} 
\begin{lstlisting}[mathescape=true, basicstyle=\footnotesize\ttfamily]
Theorem-1 $\equiv$ 
CND $\implies (\forall T \in \mathcal{T}: (\exists i \in \mathbb{N}, p \in \mathbb{P}, b \in \mathcal{B}: \L(p,b,\varsigma(\rho(T,i)))))$
\end{lstlisting}

\begin{lemma}
\label{l1}
In a fair transition path, if eventually a nonfaulty proposer $p$ becomes ready to propose a proposal number $b$, that satisfies P1 and P2, to a quorum $Q$ whose members are all nonfaulty, then eventually $p$ will receive promises from all members of $Q$ for $b$.
\end{lemma} 
\begin{lstlisting}[mathescape=true, basicstyle=\footnotesize\ttfamily]
Lemma-1 $\equiv$ 
$\forall T \in \mathcal{T}, i \in \mathbb{N}, p \in \mathbb{P}, b \in \mathcal{B}, Q \in \mathcal{Q}:$ 
	$(\dj(p) \ \land \ \th(b)\ \land \ (\forall a \in Q: (\dj(a) \ \land \ \Re(\rho(T,i), p, \mathfrak{s}(p, \langle p,a,1a,b,\bar{v} \rangle ))) ) )$
														$ \implies (\exists j \in \mathbb{N} : (j \geq i) \ \land \ \phi(p, b, Q , \varsigma(\rho(T,j))) )$
\end{lstlisting}

\begin{lemma}
\label{l2}
In a fair transition path, if eventually a nonfaulty proposer $p$ receives promises for a proposal number $b$, that satisfies P1 and P2, from a quorum $Q$ whose members are all nonfaulty, then eventually $p$ will receive votes from all members of $Q$ for $b$.
\end{lemma} 

\begin{lstlisting}[mathescape=true, basicstyle=\footnotesize\ttfamily]
Lemma-2 $\equiv$ 
$\forall T \in \mathcal{T}, i \in \mathbb{N}, p \in \mathbb{P}, b \in \mathcal{B}, Q \in \mathcal{Q}:$ 
	$(\dj(p) \ \land \ \th(b)\ \ \land \ \phi(p, Q , \varsigma(\rho(T,i))) \ \land \ (\forall a \in Q: \dj(a) ))$
														$\implies (\exists j \in \mathbb{N}: (j \geq i) \ \land \ \Phi(p, b ,Q, \varsigma(\rho(T,j))) )$
\end{lstlisting}

Given below are the proof sketches of Theorem~\ref{t1}, Lemma~\ref{l1}, and Lemma~\ref{l2}: 

\noindent\textbf{Theorem~\ref{t1} \textit{Proof Sketch}} -  
\begin{enumerate}[label = (\arabic*)]
\item By Lemma~\ref{l1} and \texttt{CND}, some nonfaulty proposer $p$ will eventually receive promises from some quorum $Q$, whose members are all nonfaulty, for some proposal number $b$ that satisfies \emph{P1} and \emph{P2}.
\item By Lemma~\ref{l2}, $p$ will eventually receive votes from $Q$ for $b$ and learn that $b$ has been chosen.  
\\
$\square$
\end{enumerate}

\noindent\textbf{Lemma~\ref{l1} \textit{Proof Sketch}} - 
\begin{enumerate}[label = (\arabic*)]
\item By \mytexttt{Prp-NF-Axm}, \mytexttt{NF-IOE-Axm}, and \mytexttt{F-Snd-Axm} \emph{prepare} messages from $p$ will eventually be sent to all members of $Q$.

\item By \mytexttt{Acc-NF-Axm}, \mytexttt{NF-IOE-Axm}, \mytexttt{F-Rcv-Axm} all members of $Q$ will eventually receive the \emph{prepare} messages.

\item By \mytexttt{P1-P2-Def}, \mytexttt{Snd-1b-Rul}, and \mytexttt{Acc-NF-Axm}, each member of $Q$ will eventually be ready to send \emph{promise} messages to $p$.

\item By \mytexttt{Acc-NF-Axm}, \mytexttt{NF-IOE-Axm}, and \mytexttt{F-Snd-Axm} the \emph{promise} messages from each member of $Q$ will eventually be sent.

\item By \mytexttt{Prp-NF-Axm}, \mytexttt{F-Rcv-Axm}, and \mytexttt{NF-IOE-Axm} $p$ will eventually receive the \emph{promise} messages from all members of $Q$.\\
$\square$
\end{enumerate}

\noindent\textbf{Lemma~\ref{l2} \textit{Proof Sketch}} -  
\begin{enumerate}[label = (\arabic*)]
\item By \mytexttt{Snd-2a-Rul}, and \texttt{Prp-NF-Axm}, $p$ will eventually be ready to send \emph{accept} messages to all members of $Q$ with proposal number $b$.

\item By \mytexttt{Prp-NF-Axm}, \mytexttt{NF-IOE-Axm}, and \mytexttt{F-Snd-Axm}, \emph{accept} messages from $p$ will eventually be sent to all members of $Q$.

\item By  \mytexttt{Acc-NF-Axm}, \mytexttt{NF-IOE-Axm}, \mytexttt{F-Rcv-Axm} all members of $Q$ will eventually receive the \emph{accept} messages.

\item By \mytexttt{P1-P2-Def}, \mytexttt{Snd-2b-Rul}, and \mytexttt{Acc-NF-Axm}, each member of $Q$ will eventually be ready to send \emph{voted} messages to $p$.

\item By \mytexttt{Acc-NF-Axm}, \mytexttt{NF-IOE-Axm}, and \mytexttt{F-Snd-Axm} the \emph{voted} messages from each member of $Q$ will eventually be sent

\item By \mytexttt{Prp-NF-Axm}, \mytexttt{F-Rcv-Axm}, and \mytexttt{NF-IOE-Axm} $p$ will eventually receive the \emph{voted} messages from all members of $Q$.\\
$\square$
\end{enumerate}

We have formalized all the theory and proof sketches presented in this section using Athena. 
The proofs of Theorem~\ref{t1}, Lemma~\ref{l1}, and Lemma~\ref{l2} have been mechanically verified for correctness.
The high-level structures of the proofs were developed in a hierarchical manner consisting of well-connected steps.
The SPASS\cite{weidenbach2009spass} automatic theorem prover was then guided with appropriate premises for mechanically verifying each step (more details can be found in the appendices at the end of this report).
We have made extensive use of Athena's existing library of natural number theory for reasoning about indexed points in transition paths.
The complete proof consists of about 6000 lines of Athena code\footnote{Complete Athena code available at http://wcl.cs.rpi.edu/pilots/fvcafp}.

%% file: related.tex
\section{Related Work}\label{related}

Prior work on verification of properties of Synod-related protocols exists in the literature. 
Prisco~\emph{et~al.}~\cite{de2000revisiting} present a detailed, formal description and a rigorous hand-written proof of safety for Paxos along with an analysis of time performance and fault tolerance. 
Chand~\emph{et~al.}~\cite{chand2016formal} provide a specification of Multi-Paxos in TLA$^{\text{+}}$~\cite{lamport2002specifying} and use TLAPS~\cite{chaudhuri2010verifying} to prove its safety.
Padon \emph{et al.}~\cite{padon2017paxos} have verified the safety property for Paxos, \emph{Vertical Paxos}~\cite{lamport2009vertical}, \emph{Fast Paxos}~\cite{lamport2006fast}, and \emph{Stoppable Paxos}~\cite{malkhi2008stoppable} using deductive verification.
K{\"u}fner~\emph{et~al.}~\cite{kufner2012formal} provide a methodology to develop machine-checkable proofs of fault-tolerant round-based distributed systems and verify the safety property for Paxos.  
Schiper \emph{et~al.}~\cite{schiper2014developing} have formally verified the safety property of a Paxos-based totally ordered broadcast protocol using EventML~\cite{bickford2012logic} and the Nuprl~\cite{naumov2001hol} proof assistant.
Howard~\emph{et~al.}~\cite{howard2016flexible} have presented \emph{Flexible Paxos} by introducing flexible quorums for Paxos and have model checked its safety property using the TLC model checker~\cite{lamport2005tlc}.
Rahli \emph{et~al.}~\cite{rahli2015formal,rahli2017eventml} have used EventML and Nuprl to formally verify the safety of an implementation of Multi-Paxos.
Attiya~\emph{et~al.}~\cite{attiya1994bounds} provide bounds on the time to reach progress in consensus by assuming a synchronous model with known bounds on message delivery and processing time of non-faulty processes.
Keidar~\emph{et~al.}~\cite{keidar2003open} consider a partial synchrony model with known bounds on processing times and message delays and use it to guarantee progress in a consensus algorithm when the bounds hold.
Malkhi~\emph{et~al.}~\cite{malkhi2008stoppable} introduce Stoppable Paxos, a variant of Paxos for implementing a stoppable state machine and provide an informal proof of safety and progress for Stoppable Paxos with a unique leader.
McMillan~\emph{et~al.}~\cite{mcmillan2018deductive} machine-check and verify the proofs of safety and progress properties of Stoppable Paxos \cite{malkhi2008stoppable} using Ivy\cite{padon2016ivy}.
Dragoi~\emph{et~al.}~\cite{druagoi2016psync} introduce PSync, a language that allows writing, execution, and verification of high-level implementations of fault-tolerant systems, and use it to verify the safety and progress properties of \emph{LastVoting}~\cite{charron2009heard}.
LastVoting is an adaptation of Paxos in the \emph{Heard-Of model}~\cite{charron2009heard} that guarantees progress under the assumption of a single leader.
A machine-checked proof of safety and progress of LastVoting also appeared in \cite{debrat2012verifying}.
Hawblitzel~\emph{et~al.}~\cite{hawblitzel2015ironfleet,hawblitzel2017ironfleet} introduce a framework for designing provably correct distributed algorithms called IronFleet and use it to prove the safety and progress properties for a Multi-Paxos implementation called IronRSL.
They embed TLA$^{\text{+}}$ specifications in Dafny~\cite{leino2010dafny} for formally verifying IronRSL. 
Their proof of progress relies on the assumption that eventually, all messages will arrive within a maximum network delay and leader election will succeed.

All of the aforementioned work has either analyzed the safety property or both the safety and progress properties of Synod-related protocols. 
Where progress has been verified, the authors have either assumed a unique leader, synchrony, or both.
Our work improves upon existing work by identifying a set of asynchronous conditions under which the fundamental Synod consensus protocol can make eventual progress in the absence of a unique leader, and providing the first mechanically verified proof of eventual progress in Synod.

%% file: conc.tex
\section{Conclusion}\label{conc}

We have identified a set of sufficient conditions under which the Synod protocol can make progress, in asynchronous communication settings and in the absence of a unique leader.
Leader election itself being a consensus problem, our conditions generalize Paxos' progress conditions by eliminating their cyclic reliance on consensus. 
Consequently, our weaker assumptions do not impose a communication bottleneck or proposal restrictions.
We have introduced a failure-aware actor model (FAM) to reason about communication failures in actors.
Using this reasoning framework we have formally demonstrated that eventual progress can be guaranteed in Synod under the identified conditions. 
Finally, we have used Athena to develop the first machine-checked proof of progress in Synod.

It is important to note that a guarantee of eventual progress only states that consensus will be achieved, but does not provide any bound on the amount of time that may be required for the same.
Since air traffic data usually has a short useful lifetime and aircraft have limited time to remain airborne, a guarantee of eventual progress \emph{alone} is insufficient for UAM applications.
To be useful, a progress guarantee should have some associated time bounds that the aircraft can use to make important decisions,
\emph{e.g.}, if there is a guarantee that consensus will take at least 5 seconds, then a candidate can decide to only compute flight plans that start after 5 seconds.
Nevertheless, we see this work as a valuable exercise in perceiving the nuances involved in guaranteeing eventual consensus in the presence of multiple unrestricted proposers. 
This is important because a guarantee of eventual progress is a necessary precondition for providing a guarantee of timely progress that can be directly applicable for UAM. 

A potential direction of future work would be to investigate formal proofs of probabilistic guarantees of timely progress by using data-driven statistical results.
Such properties may be provided by using statistical observations about message transmission and processing delays, which cannot be deterministically predicted in asynchronous conditions but can be observed at run-time.
Another potential direction of work would be to model message loss in FAM by introducing additional meta-level transitions.
This would allow us to weaken the conditions further by requiring the guaranteed delivery of only a subset of messages, thereby weakening the current fairness assumptions of FAM.
To avoid livelocks, it may also suffice to replace predicate fairness with a weaker assumption that \emph{infinitely often enabled finite transition sequences must eventually occur}.

%% file: acknowledge.tex
\section*{Acknowledgement}
This research was partially supported by the National Science Foundation (NSF), Grant No. -- CNS-1816307 and the Air Force Office of Scientific Research (AFOSR), DDDAS Grant No. -- FA9550-19-1-0054.
The authors would like to express their gratitude to Elkin Cruz-Camacho, Dan Plyukhin, and the anonymous reviewers of NFM 2021 for their helpful comments on improving the manuscript.

%% file: appendices.tex
\newpage
\begin{appendices}
\chapter{Using SPASS with Athena}

Athena is integrated with automated theorem provers (ATP) like Vampire and SPASS that can be invoked seamlessly like primitive procedures.
In the case of our proof of eventual progress for Synod, we had divided the proof into a series of hierarchical proof obligations that were well-connected by trivial operations like \emph{modus ponens}, \emph{conjunction elimination}, \emph{case analysis}, etc.
However, the complexity of the statements involved made it cumbersome and time-consuming to write the proofs for these obvious steps using the Athena proof semantics, making the task of completing the entire proof extremely time-consuming.
For this reason, we used the SPASS theorem prover to generate the proofs of the trivial operations that connected one proof obligation to the next obligation in the proof chain. 
We present such an example below to demonstrate how SPASS was used in our proofs.

Let us consider the example of \mytexttt{IOE->Fair-Rcv-Causes-Theorem}, which is an intermediate theorem that states that for a nonfaulty actor, if a receive transition is enabled for a message at a time, then the message will eventually be in the set of unresponded messages in the actor's local state.

\begin{lstlisting}[style=athena-list]
define Fair-Rcv-Causes-Theorem :=
(forall x m T i .
	(nonfaulty x) 
		==>
		(
	        ((inMSet m (mu (config (rho T i)))) 
	        & (ready-to (rho T i) x (receive x m)))  
            	==>
	            	(exists j . (i N.<= j) 
        	    				& (inMSet m (amu (als (config (rho T j)) x))))  						
		)

 )	
\end{lstlisting}

To prove \mytexttt{IOE->Fair-Rcv-Causes-Theorem}, we use four facts which have been previously added to the assumption base: \mytexttt{S3}, which states that every natural number is either less than or equal to its successor; \mytexttt{transitive}, which states the transitive rule corresponding to the \emph{less-than or equal to} relationship of natural numbers; 
\mytexttt{IOE->Fair-Rcv-Theorem}, which states that for a nonfaulty actor, if a receive transition is enabled at some time, then it will eventually occur;
and \mytexttt{rcv-affect}, which simply states that if a receive transition happens, then the corresponding message exists in the set of unresponded messages in the recipient's new local state.   

\begin{lstlisting}[style=athena-list]
define S3 := (forall n . n <= S n)

define transitive := (forall x y z . x <= y & y <= z ==> x <= z)

define rcv-affect :=
(forall x m T i j .
        (= (rho T i) (then (rho T j) (receive x m))) 
    	==>
        	  (inMSet m (amu (als (config (rho T i)) x))) 
 )

define IOE->Fair-Rcv-Theorem :=
(forall x m T i .
	(nonfaulty x) 
		==>
		(
	        ((inMSet m (mu (config (rho T i)))) 
	        & (ready-to (rho T i) x (receive x m)))  
            	==>
	            	(exists j . (i N.<= j) 
        	    				& (= (rho T (S j)) (then (rho T j) (receive x m))) ) 					
		)
 )
\end{lstlisting}

We can see that to obtain \mytexttt{IOE->Fair-Rcv-Causes-Theorem}, from the four facts, it is a simple application of \mytexttt{rcv-affect} to \mytexttt{IOE->Fair-Rcv-Theorem}, to show that the message will be received and added to the local state, followed by applications of \mytexttt{S3} and \mytexttt{transitive}.
There are many ways to develop the proof interactively in Athena, but due to the complexity of the statements of \mytexttt{IOE->Fair-Rcv-Theorem-step1}, it becomes a very convoluted task to develop the proof using Athena proof semantics, even though the logic is straightforward.
Given below is one possible way to develop the proof interactively in Athena.

\begin{lstlisting}[style=athena-list]
conclude 
(forall x m T i .
	(nonfaulty x) 
		==>
		(
	        ((inMSet m (mu (config (rho T i)))) 
	        & (ready-to (rho T i) x (receive x m)))  
            	==>
	            	(exists j . (i N.<= j) 
        	    				& (inMSet m (amu (als (config (rho T j)) x))))  						
		)

 )
pick-any x:Actor
pick-any m:Message
pick-any T:TP 
pick-any i:N    
assume 
		(nonfaulty x)
assume 
		( (inMSet m (mu (config (rho T i)))) 
	    & (ready-to (rho T i) x (receive x m)))    
let{
	IOE->FAIR-Rcv := (!uspec (!uspec (!uspec (!uspec IOE->Fair-Rcv-Theorem x) m) T) i);
	mp1 := (!mp 
    		    IOE->FAIR-Rcv 
    		   (nonfaulty x)
		   );
	exst-j := (!mp 
    		       mp1 
					( (inMSet m (mu (config (rho T i)))) 
		    	    & (ready-to (rho T i) x (receive x m)))    
    	   );
	i<=Si := (!uspec N.Less=.S3 i);
	trueforSj := conclude (exists z . 
		                      (i N.<= z) 
	            	        & (inMSet m (amu (als (config (rho T z)) x))))
				 pick-witness j for exst-j
				 let{
				 	exst-j-sttmnt := ( (i N.<= j) 
	            	    		     & (= (rho T (S j)) (then (rho T j) (receive x m))));
				 	Sj := (S j);
				 	j<=Sj := (!uspec N.Less=.S3 j);
				 	i<=j := (!left-and exst-j-sttmnt);
				 	trnstv-sttmnt := (!uspec (!uspec (!uspec N.Less=.transitive i) j) Sj);
				 	i<=j&j<=Sj := (!both i<=j j<=Sj);
				 	rcv-efctSj := (!uspec (!uspec (!uspec (!uspec (!uspec rcv-affect x) m) T) Sj) j);
				 	i<=Sj := (!mp 
				 				trnstv-sttmnt
				 				i<=j&j<=Sj
				 		      );
				 	i<=Sj&Sjthen := (!both i<=Sj (!right-and exst-j-sttmnt));
				 	Sjinmset := conclude (inMSet m (amu (als (config (rho T (S j))) x)))
				 				let{
    				 				implies := (!chain [ (= (rho T (S j)) (then (rho T j) (receive x m)))
    				 								==>  (inMSet m (amu (als (config (rho T (S j))) x)))  [rcv-efctSj]
    				 					    			])	
				 				}	
				 				(!mp implies (!right-and exst-j-sttmnt));
				 	i<=Sj&Sjinmset := (!both i<=Sj Sjinmset)			
				 }
				 (!egen 
	            	(exists j . (i N.<= j) 
        	    				& (inMSet m (amu (als (config (rho T j)) x))))
        	    	(S j)			    				 		
				 	)			 		
	}
	(!claim trueforSj)
\end{lstlisting}

Developing such proofs in a completely interactive manner gives confidence about the logical soundness of the proof steps involved. 
However, in significantly large projects, like our verification of eventual progress in Synod, which involves handling a significant number of such proof obligations, the task can easily become arduous and time-consuming.
Therefore, in our verification of eventual progress in Synod, we have used SPASS to aid in the proof development effort.
SPASS can be automatically invoked within Athena by using a statement of the form \texttt{(!prove p L)} where \mytexttt{p} is the conclusion that we want to derive and \mytexttt{L} is a list of facts that are already in the assumption base which we want SPASS to use as a premise for the proof.
If SPASS can successfully derive the required conclusion, then the statement is added as a theorem into the assumption base.
In many cases though, SPASS may be unable to derive the conclusion from the premises passed to it, causing it to timeout and exit. 
Also, in many cases, SPASS confirms before timing out that the desired conclusion cannot be derived from the premises.

In our verification of progress in Synod, we have divided complicated proof obligations into smaller obligations that are well-connected by trivial steps and then used SPASS to prove each intermediate step. 
Below, we show an example of this approach by using the proof of \mytexttt{IOE->Fair-Rcv-Causes-Theorem} which has been broken down into two obvious intermediate steps.
Each step was proven by calling SPASS with the required premises.

\begin{lstlisting}[style=athena-list]
define Fair-Rcv-Causes-Theorem-step1 :=
(forall x m T i .
	(nonfaulty x) 
		==>
		(
	        ((inMSet m (mu (config (rho T i)))) 
	        & (ready-to (rho T i) x (receive x m)))  
            	==>
	            	(exists j k . (i N.<= j)
	            				& (j N.<= k) 
	            				& (= (rho T k) (then (rho T j) (receive x m)))  						
        	    				& (= (rho T (S j)) (then (rho T j) (receive x m)))) 						
		)
 )	

(!prove Fair-Rcv-Causes-Theorem-step1 [IOE->Fair-Rcv-Theorem N.Less=.S3])

define Fair-Rcv-Causes-Theorem-step2 :=
(forall x m T i .
	(nonfaulty x) 
		==>
		(
	        ((inMSet m (mu (config (rho T i)))) 
	        & (ready-to (rho T i) x (receive x m)))  
            	==>
	            	(exists j k . (i N.<= j)
	            				& (j N.<= k) 
	            				& (= (rho T k) (then (rho T j) (receive x m)))  						
        	    				& (= (rho T (S j)) (then (rho T j) (receive x m)))
        	    				& (inMSet m (amu (als (config (rho T k)) x)))) 						
		)
 )	

(!prove Fair-Rcv-Causes-Theorem-step2 [Fair-Rcv-Causes-Theorem-step1 rcv-affect])

# Finally, proving Fair-Rcv-Causes-Theorem
(!prove Fair-Rcv-Causes-Theorem [Fair-Rcv-Causes-Theorem-step2 N.Less=.transitive])
\end{lstlisting}

Athena is \emph{sound}, \emph{i.e.}, if a deduction $D$ is evaluated with respect to an assumption base $\beta$ and produces a sentence $p$, then Athena provides the guarantee that $p$ is a logical consequence of $\beta$.  
Athena can also detect \emph{ill-sorted} expressions in proofs and report them, thereby removing the possibilities of ill-formed expressions in the proofs and assumption base.
However, we are aware that automated theorem provers like SPASS are like \emph{black-boxes} whose internal operations cannot be analyzed.
If there is an inconsistency in the premise passed to an ATP, the ATP may derive \texttt{false} from the premise, enabling it to erroneously deriving any desired conclusion.
Therefore, when using ATPs, one must make sure that inconsistent statements are not passed as premises for deriving conclusions.
For this reason, we have taken the following steps to ensure that we can have a high degree of confidence in the proofs that we have developed using SPASS:
\begin{itemize}
    \item We have checked that the facts in our assumption base are not inconsistent by making sure that they conform to the informal proofs that we have developed.
    \item We have checked that SPASS cannot generate \texttt{false} from the axioms in our assumption base.
    \item We have kept the intermediate obligations in our proof hierarchy simple enough so that they can be obtained from their preceding steps by basic operations like \emph{modus ponens}, \emph{conjunction elimination}, etc. 
    In this way, we have ensured that we know what SPASS is supposed to do before we pass the obligations to it instead of using SPASS like a black-box.
    \item Every time we have used SPASS to prove some obligation from a premise \texttt{L}, we have checked the consistency of \texttt{L} by using the command \texttt{(!prove false  L)} to ensure that SPASS cannot use any inconsistencies in the premise to erroneously prove the required obligation.
    In all cases, when we have tried to prove \texttt{false} from a premise we wanted to use for some obligation, SPASS almost instantly failed to prove \texttt{false} without timing out, giving us high confidence that the premise was consistent.

\end{itemize}
Given the soundness guarantees of Athena, its built-in sort-detection feature, and the steps we have taken to avoid inconsistencies in our proofs, we have established a reasonable degree of confidence in the soundness of our approach. 


\chapter{Athena Code for the Proofs}

\section{Relating the Athena code to the formal theory}

\begin{table}[htp]
\centering

\begin{adjustbox}{width=\textwidth}

\begin{tabular}{|c|c|c|}
\hline
\textbf{Symbol} & \textbf{Description} & \textbf{Athena Counterpart} \\ \hline
$\varsigma$     &  Get last configuration            &  $\texttt{config}$      \\ \hline
$\rho$    &    Get transition path up to index         &     $\texttt{rho}$     \\ \hline
$\top$      &  Transition path constructor         &   $\texttt{then}$       \\ \hline
$\sigma$      &  Choose a value to propose based on configuration       & $\texttt{decide-value}$       \\ \hline
$\mathfrak{s}$      &  Construct a \texttt{snd} transition step         & $\texttt{send}$       \\ \hline
$\mathfrak{r}$      &  Construct a \texttt{rcv} transition step         & $\texttt{receive}$       \\ \hline
$\mathfrak{m}$      &  Get set of messages ``en route''           & $\texttt{mu}$       \\ \hline
$\mathfrak{a}$      &  Actor is in $\alpha$    & $\texttt{available}$       \\ \hline
$\Re$      &  Actor is ready for a step               & $\texttt{ready-to}$       \\ \hline
$\phi$      &  Proposer has promises from a quorum              & $\texttt{has-promises}$       \\ \hline
$\Phi$      &  Proposer has votes from a quorum                 & $\texttt{has-votes}$       \\ \hline
$\dj$      &  Actor is nonfaulty       & $\texttt{nonfaulty}$       \\ \hline
$\th$      &  Proposal number satisfies \emph{P1} and \emph{P2}        & $\texttt{P1+P2-True}$       \\ \hline
$\L$      &  Proposer has learned of successful consensus            & $\texttt{learn}$       \\ \hline
\end{tabular}

\end{adjustbox}

\caption{Athena counterparts to the symbols in our formal representation.}
\end{table}

Table B.1 connects the symbols used to express the formal theory presented in Section~\ref{proof} to the actual code constructs used in Athena.
This will be useful for readers attempting to comprehend the code.

\section{The Athena codes}
\begin{lstlisting}[style=athena-list]
#****************************************************************************
# The symbols needed to express the Synod algorithm using the actor model 
#****************************************************************************

module Synod{

	#-- Actors
	domain Actor

	#-- Values that can be proposed
	domain Value
	declare nilV : Value
	declare equal-val : (Value) [Value Value] -> Value [=]

	#-- Ballot (not directly using natural numbers)
	domain Ballot 

	#-- Syntactic Sugar Predicate that returns true if the ballot satisfies P1+P2-True properties for progress
	declare P1+P2-True: (Ballot) [Ballot] -> Boolean	

	#-- Some basic relations over ballots (similar to natural numbers)
	declare less-than : (Ballot) [Ballot Ballot] -> Boolean [<]
	declare less-than-equal : (Ballot) [Ballot Ballot] -> Boolean [=<]
	declare greater-than : (Ballot) [Ballot Ballot] -> Boolean [>]
	declare greater-than-equal : (Ballot) [Ballot Ballot] -> Boolean [>=]
	declare equal : (Ballot) [Ballot Ballot] -> Boolean [=]

	#-- Predicates over the set of actors (creates two subests, one each for proposers and acceptors)
	declare prp: (Actor) [Actor] -> Boolean
	declare acc: (Actor) [Actor] -> Boolean

	#-- Syntactic Sugar Predicate that returns true if an actor satisfies the required nonfaulty behavior for progress 
	declare nonfaulty: (Actor) [Actor] -> Boolean

	#-- Set od all Non-empty Quorums
	datatype Quorum := Actor 
					| (consQ Quorum Actor)
	declare inQ : (Actor, Quorum) [Actor Quorum] -> Boolean

	#-- Message types
	domain Type

	#-- Messages
	datatype Message := (consM frm:Actor to:Actor typ:Type bal:Ballot val:Value)

	#-- Get functions for messages
	declare msgSndr: (Message) [Message] -> Actor
	declare msgRcpnt: (Message) [Message] -> Actor
	declare msgTyp: (Message) [Message] -> Actor
	declare msgBal: (Message) [Message] -> Actor
	declare msgVal: (Message) [Message] -> Actor

	#-- Set of messages
	datatype MessageSet := nilMSet 
						| (consMSet MessageSet Message)

	declare inMSet : (Message, MessageSet) [Message MessageSet] -> Boolean [in]

	#-- Actor Local State 
	datatype ALS := (consALS ms:MessageSet bal:Ballot bal':Ballot val':Value lck:Boolean)

	#-- Functions that return values from actor local state
	declare amu: (ALS) [ALS]-> MessageSet # set of messages received but not responded to yet 
	declare beta: (ALS) [ALS]-> Ballot    # highest ballot seen
	declare beta_cap: (ALS) [ALS]-> Ballot # highest ballot accepted
	declare value: (ALS) [ALS]-> Value    # value corresponding to beta_cap
	declare lock: (ALS) [ALS]-> Boolean  #
	declare init: (ALS) [ALS]-> Boolean # to store if it has initiated any ballot yet (used for livelock proof)
	declare has-promises : (ALS, Ballot, Quorum) [ALS Ballot Quorum]-> Boolean # returns true if it has received 1b from a quorum
	declare decide-value : (ALS) [ALS]-> Value # Selects the value to be proposed depending on values sent by quorum. Otherwise returns some new value.
	declare has-votes : (ALS, Ballot, Quorum) [ALS Ballot Quorum]-> Boolean #returns true if it has received 2b from a quorum
	declare sent2A : (ALS, Quorum, Ballot, Value) [ALS Ballot Value Quorum N]-> Boolean #proposer has sent 2A mssages to a quorum

	#-- Set of actors 
	datatype ActorSet := nilASet 
					| (consASet ActorSet Actor)

	declare inASet : (Actor, ActorSet) [Actor ActorSet] -> Boolean 

	#-- Domain Configuration
	domain Configuration

	#-- Functions that return the set of in-transit messages an actors local states in a configuration 
	declare mu : (Configuration) [Configuration] -> MessageSet # gets the multiset of messages "en route" 
	declare als: (Configuration, Actor) [Configuration Actor] -> ALS # gets actor local state from configuration 
	declare alpha: (Configuration) [Configuration] -> ActorSet # gets the mapping of available actors
	declare alpha_bar: (Configuration) [Configuration] -> ActorSet # gets the mapping of failed actors

	#-- Datatype of transition steps (create not needed for Synod proof)
	datatype Step := (receive a:Actor m:Message) 
	               | (send a:Actor m:Message) 
	               | (create old:Actor new:Actor)
	               | (stop a:Actor )
	               | (start a:Actor ) # "begin" keyword taken in Athena

	#-- Datatype of transition paths borrowed from Musser
	datatype TP := Initial 
	                     | (then TP Step)

	declare config :  [TP ] -> Configuration # last configuration in a given path
	declare ready-to : [TP  Actor Step] -> Boolean # if actor is ready to perform some step in the last configuration in a given path
	declare available :  [TP  Actor] -> Boolean # if actor is available in last configuration in a given path

	#-- The following creates a sub-transition path upto the given index
	#-- used for accessing indexed positions in transition paths
	declare rho :  [TP  N] -> TP

	#-- Predicate that checks if a proposer has obtained 
	#-- votes from a quorum for a ballot (learnt of successful consensus)
	declare learn: (Actor, Configuration, Ballot) [Actor Configuration Ballot] -> Boolean

	#-- Declaring all constants and variables used in the rest of the proofs 
	#-- There are four types of messages
	declare 1a, 1b, 2a, 2b, failT : Type     

	# -- Set of variables used throughout module Synod
	define [p a x Q b bH b2 b3 v v2 v3 m m2 ms1 ms2 c s T i j k l y z w e f t u p2 a2 typ sender' recipient'] :=
																					 	        [   ?p:Actor 
																					 	        	?a:Actor 
																					 	        	?x:Actor 
																					 	        	?Q:Quorum 
																					 	        	?b:Ballot 
																					 	        	?bH:Ballot 
																					 	        	?b2:Ballot
																					 	        	?b3:Ballot
																					   	        	?v:Value
																					   	        	?v2:Value
																					   	        	?v3:Value 
																					   	        	?m:Message 
																					   	        	?m2:Message
																					   	        	?ms1:MessageSet 
																					   	        	?ms2:MessageSet 
																					   	        	?c:Configuration 
																					   	        	?s:Step
																					   	        	?T:TP 
																					   	        	?i:N 
																					   	        	?j:N 
																					   	        	?k:N
																					   	        	?l:N
																					   	        	?y:N
																					   	        	?z:N
																					   	        	?w:N
																					   	        	?e:N
																					   	        	?f:N
																					   	        	?t:N
																					   	        	?u:N
																					   	        	?p2:Actor
																					   	        	?a2:Actor
																					   	        	?typ:Type
																					   	        	?sender':Actor
																					   	        	?recipient':Actor  ]

}#close module Synod	
\end{lstlisting}

\begin{lstlisting}[style=athena-list]
#****************************************************************************
# Constraints/rules defining the general system model following the actor model and paxos 
#****************************************************************************
load "lib/main/nat-less"
load "symbols.ath"

extend-module Synod{

#----------------------------------------------------------------------------
#-- Availability syntactic sugar (unused for progress)

	#-- An actor is available iff it it in the alpha set of the configuration
	define availability-axiom :=
		(forall T i x .
			(available (rho T i) x)
				<==> 
					(inASet x (alpha (config (rho T i))))
		)

#----------------------------------------------------------------------------
#-- Ready-to fair-axioms (unused for progress)

	#-- An actor can only be ready for a transition if it is available
	define ready-to-axiom :=
		(forall T i x s .
			(ready-to (rho T i) x s)
				==> 
					(available (rho T i) x)
		)

#----------------------------------------------------------------------------
#-- Transition Step Syntactic Sugar 

	#-- If a receive trasition happens at an indexed position, the message exists 
	#-- in the set of unresponded messages in the local state of the actor at that position 
	define rcv-affect :=
			(forall x m T i j .
			        (= (rho T i) (then (rho T j) (receive x m))) 
	            	==>
		            	  (inMSet m (amu (als (config (rho T i)) x))) 
             )

	#-- If a Send trasition happens at an indexed position, the message exists  
	#-- in the multiset at that position 
	define snd-affect :=
			(forall x m T i j .
			        (= (rho T i) (then (rho T j) (send x m)))
	            	==>
		            	(inMSet m (mu (config (rho T i))) )
             )

	define model-step-axioms := [ rcv-affect
								  snd-affect
							    ]

#----------------------------------------------------------------------------
#-- Actor Fairness theory 

	#-- The fair receive and fair send assumptions of the actor model
	define fair-rcv :=
			(forall x m T i .
			        ((inMSet m (mu (config (rho T i)))) 
			        & (ready-to (rho T i) x (receive x m)))  
		            	==>
		            	(
		            		  (exists j . (i N.<= j) 
		            		   			& (= (rho T (S j)) (then (rho T j) (receive x m))) )
		            	    | (exists k . (i N.< k)
		            	    			& (forall j .
		            	    			   (k N.<= j)
		            	    			     ==>
		            	    			     	(~(
				            	    			   (inMSet m (mu (config (rho T j)))) 
				            	    			 & (ready-to (rho T j) x (receive x m))		            	    			     		
		            	    			     	 ))
		            	    			  )

		            	       )		
		            	) 
             )

	define fair-snd :=
			(forall x m T i .
			        (ready-to (rho T i) x (send x m))  
		            	==>
		            	(
		            		  (exists j . (i N.<= j) 
		            		   			& (= (rho T (S j)) (then (rho T j) (send x m))) )
		            	    | (exists k . (i N.< k)
		            	    			& (forall j .
		            	    			   (k N.<= j)
		            	    			     ==>
		            	    			     	(~(ready-to (rho T j) x (send x m)))
		            	    			  )

		            	       )		
		            	) 
             )

	define model-fair-axioms := [ fair-rcv 
							      fair-snd
							    ]

#----------------------------------------------------------------------------
#-- Message Persistance Property

	# Messages in the multiset persist until received (can be proven, but currently taking as an axiom)
	define msg-persists-until-read :=
			(forall T i a p typ b v .
					(inMSet (consM p a typ b v) (mu (config (rho T i))))			
					==>
						(
							  (forall j .
							        (i N.<= j) 
							  		      ==> 
							  			     (inMSet (consM p a typ b v) (mu (config (rho T j)))) 
							  )

							| (exists k . (   (i N.<= k) 
		            	    				& (= (rho T (S k)) (then (rho T k) (receive a (consM p a typ b v))))
		            	    			  )
							  )
						)	
			)

}#close module
\end{lstlisting}

\begin{lstlisting}[style=athena-list]
#****************************************************************************
# Constraints/rules defining the Proposer and Acceptor behavior for Synod
#****************************************************************************
load "lib/main/nat-less"
load "symbols.ath"

extend-module Synod{

#----------------------------------------------------------------------------
#-- Quorum theory 

	#-- Non-empty quorums (Unused for progress)
	define synod-non-empty-Q :=
			(forall Q . (exists a . (inQ a Q)))

	#-- All members of a quorum are acceptors
	define synod-only-acc-inQ :=
			(forall Q a . 
				((inQ a Q) ==> (acc a))
			)

	define synod-quorum-axioms := [ synod-non-empty-Q
									synod-only-acc-inQ 
								  ]

#----------------------------------------------------------------------------
#-- Definitions of the has-promises, has-votes, and learn predicates

	define synod-has-promises-def-axiom :=
        (forall p T b i Q . 
				(forall a . (inQ a Q) ==> (exists v . (inMSet (consM a p 1b b v) (amu (als (config (rho T i)) p))) ))
	    			<==> 
	    			    (has-promises (als (config (rho T i)) p) b  Q) 

        )

	define synod-has-votes-def-axiom :=
        (forall p T b i Q . 
				(forall a . (inQ a Q) ==> (exists v . (inMSet (consM a p 2b b v) (amu (als (config (rho T i)) p))) )) # To ensure safety, v should be same for all a in Q
	    			<==> 
	    			    (has-votes (als (config (rho T i)) p) b  Q) 

        )

	#-- A proposer learns a ballot has been chosen if there is a time when its local
	#-- state has 2b messages for that ballot from a quorum of acceptors	
	define synod-learn-def-axiom :=
	(forall T i p b Q .
	    (
  			(has-votes (als (config (rho T i)) p) b  Q)
				<==>
					(learn p (config (rho T i)) b)
	  	)					  
    )

#----------------------------------------------------------------------------
#-- Message Sending Behavior of Proposers and Acceptors depending on local state

	#-- If an actor is a proposer, then if it has received 1b messages from every acceptor in a Quorum Q 
	#-- then it becomes ready to send 2a message to each acceptor in Q iff it is available

	define synod-2a-msg-rule := 
        (forall p T b i Q . 
        	(prp p) ==>
	            ( 
					(has-promises (als (config (rho T i)) p) b  Q)
	                    ==> 
	                        (
	                            (available (rho T i) p)
	                                ==>
	                                     ( (forall a .
	                                            (inQ a Q) 
	                                                ==> 
	                                                   (ready-to (rho T i) p (send p (consM p a 2a b (decide-value (als (config (rho T i)) p)) ))) ) )
	                        )
	            ) 
        )

	#-- If an actor is an acceptor and it receives an 1a message from a proposer for a ballot greater than any ballot it has seen, 
	#-- then it becomes ready to send 1b message to the proposer

	define synod-1b-msg-rule := 
		(forall p T b i a . 
			(acc a) ==>
				( 
					( (inMSet (consM p a 1a b nilV) (amu (als (config (rho T i)) a))) 
					& ((beta (als (config (rho T i)) a)) < b) )
						==> 
							(
								(available (rho T i) a)
									==>
										 (ready-to (rho T i) a (send a (consM a p 1b b (value (als (config (rho T i)) a))) ) )
							)
				) 
		)

    #-- If an actor is an acceptor and it receives an 2b message from a proposer for a ballot greater or equal to any ballot it has seen, 
	#-- then it becomes ready to send 2b message to the proposer

	define synod-2b-msg-rule := 
		(forall p T b i a v . 
			(acc a) ==>
				( 
					( (inMSet (consM p a 2a b v) (amu (als (config (rho T i)) a))) 
					& ((beta (als (config (rho T i)) a)) =< b) )
						==> 
							(
								(available (rho T i) a)
									==>
										 (ready-to (rho T i) a (send a (consM a p 2b b v) ) ) 
							)

				) 
		)

	define synod-ready-to-send-axioms := [ 
				   						  synod-1b-msg-rule
									      synod-2a-msg-rule
									      synod-2b-msg-rule
								         ]

#----------------------------------------------------------------------------
#-- Message Receiving Behavior of Proposers and Acceptors 

	define synod-ready-to-receive-general-axiom :=
		(forall T i sender' recipient' b typ v .
				  (	
					    (available (rho T i) recipient')
					  & (inMSet (consM sender' recipient' typ b v) (mu (config (rho T i))))
				  )
					<==>
						(ready-to (rho T i) recipient' (receive recipient' (consM sender' recipient' typ b v)))
					
		)

#----------------------------------------------------------------------------
#-- Some axioms that are a consequence of Stable Storage in FAL  
#-- FAL only allows failures in which stable storage of actors persists 
#-- We use an axiom that asserts that any message in the set of unresponded messages
#-- in the local state of synod actors will persist in the set until it has been responded to 

	#-- sent2A predicate is a syntactic sugar that states that 2a messages have been sent to a quorum
	define sent2A-def :=
    (forall p T b i Q v .

    	(sent2A (als (config (rho T i)) p) b v Q i) <==>
						      (forall a .
			                    (inQ a Q) 
			                    	==> 
			                    	   (exists u .
			                    	   	        (i N.<= u)
											  #& (= (rho T (S u)) (then (rho T u) (send p (consM p a 2a b v))))
											  & (inMSet (consM p a 2a b v) (mu (config (rho T u))) )	
			                    	   )
			                  )	
    )

#-- Pattern matched persistance axioms for various types of unresponded Synod messages 
	define persistent-unresponded-local-1a-msgs :=
	(forall T i a p b .
		(
		 (inMSet (consM p a 1a b nilV) (amu (als (config (rho T i)) a))) 
		 	==>
					  (forall j .
					        (i N.<= j) 
					  		      ==> 
					  			      (inMSet (consM p a 1a b nilV) (amu (als (config (rho T j)) a))) 
					  )
					| (exists k . 
							(i N.<= k)
							& (= (rho T (S k)) (then (rho T k) (send a (consM a p 1b b (value (als (config (rho T k)) a)))))) 
					  )
		)
    )

	define persistent-unresponded-local-2a-msgs :=
	(forall T i a p b v .
		(
		 (inMSet (consM p a 2a b v) (amu (als (config (rho T i)) a))) 
		 	==>
					  (forall j .
					        (i N.<= j) 
					  		      ==> 
					  			      (inMSet (consM p a 2a b v) (amu (als (config (rho T j)) a))) 
					  )
					| (exists k . 
							(i N.<= k)
							& (= (rho T (S k)) (then (rho T k) (send a (consM a p 2b b v)))) 
					  )
		)
    )    

	define persistent-unresponded-local-promises :=
	(forall T i p Q b .
		(
		 (has-promises (als (config (rho T i)) p) b Q) 
		 	==>
		 		(
					  (forall j .
					        (i N.<= j) 
					  		      ==> 
					  			      (has-promises (als (config (rho T j)) p) b Q) 
					  )
					| (exists k v . 
							  (i N.<= k)
						    & (sent2A (als (config (rho T k)) p) b v Q k)	
					  )
				)	
		)
    )

	# If a proposer P receives a particular type of message from all members of a quorum Q at different times, then
	# eventually that message will exist in P's set of unresponded messages from all members of Q. 
	# This is because P can only respond to a message after it has received a version of it from all members of a quorum. 
    define received-from-quorum-general :=
	(forall T i p b Q typ .
	    (
  			(forall a . (inQ a Q) ==> (exists j v . (i N.<= j) & (inMSet (consM a p typ b v) (amu (als (config (rho T j)) p))) ))
			==>
				(exists k . (i N.<= k) 
					      & (forall a . (inQ a Q) ==> (exists v . (inMSet (consM a p typ b v) (amu (als (config (rho T k)) p))) )) )
	  	)					  
    )

	define synod-state-axioms := [	sent2A-def
									persistent-unresponded-local-1a-msgs
									persistent-unresponded-local-2a-msgs
									persistent-unresponded-local-promises
									received-from-quorum-general
								 ]

}#close module
\end{lstlisting}

\begin{lstlisting}[style=athena-list]
#****************************************************************************
# Constraints/conditions required specifically for proving progress in Paxos 
#****************************************************************************
load "lib/main/nat-less"
load "symbols.ath"

extend-module Synod{

#----------------------------------------------------------------------------
#-- GUARANTEES IMPLIED BY THE "nonfaulty" PREDICATE  
#----------------------------------------------------------------------------

#-- The "nonfaulty" predicate basically implies that an actor will be available
#-- when the system conditions require it to perform some action, i.e., a 
#-- nonfaulty actor is available whenever it is supposed to send a message or receive
#-- a message that had been sent to it 

	#-- A nonfaulty acceptor is available if there is a message for it in the multiset
	define acceptor-nonfaulty1 :=
	(forall a .
		( (   (acc a) 
			& (nonfaulty a) 
		  )
				==>
					(forall p T b i v typ .

	  		  	       (inMSet (consM p a typ b v) (mu (config (rho T i))))
	        	  		  ==> (
	        	  		  		  (available (rho T i) a)

	    		    	  		| (exists j .
                                             (i N.< j)
	    		    	  				   & (available (rho T j) a))
	    		    	  	  )		
					)
		) 
	)

	#-- An nonfaulty acceptor is available if its local state satisfies conditions for it to send a 
	#-- 1b or 2b message 
	define acceptor-nonfaulty2 :=
	(forall a .
		( (   (acc a) 
			& (nonfaulty a) 
		  )
				==>
					(forall p T b i v typ .
					  	(  (inMSet (consM p a typ b v) (amu (als (config (rho T i)) a)))  
					  	 & (   ((beta (als (config (rho T i)) a)) < b)
                             | ((beta (als (config (rho T i)) a)) =< b)
					  	   )	
                        )
	        	  		  ==> (
	        	  		  		  (available (rho T i) a)

	    		    	  		| (exists j .
                                             (i N.< j)
	    		    	  				   & (available (rho T j) a))
	    		    	  	  )			    
					)
		) 
	)

	#-- A nonfaulty proposer is available if there is a message for it in the multiset
	define proposer-nonfaulty1 :=
	(forall p .
		( (   (prp p) 
			& (nonfaulty p) 
		  )
				==>
					(forall T i a typ b v .   
        	   		  	  (inMSet (consM a p typ b v) (mu (config (rho T i))) )
		
	        	  		  ==> (
	        	  		  		  (available (rho T i) p)

	    		    	  		| (exists j .
                                             (i N.< j)
	    		    	  				   & (available (rho T j) p))
	    		    	  	  )			    
					)
		) 
	)

	#-- A nonfaulty proposer is available if its local state satisfies conditions for it to send a 
	#-- 2a message 
	define proposer-nonfaulty2 :=
	(forall p .
		( (   (prp p) 
			& (nonfaulty p) 
		  )	
				==>
					(forall T i typ b v Q .
						(
							(forall a .
								(inQ a Q) 
									==> 
										(exists j v . (inMSet (consM a p 1b b v) (amu (als (config (rho T j)) p))) )	       
							)
	        	  		  ==> (
	        	  		  		  (available (rho T i) p)

	    		    	  		| (exists j .
                                             (i N.< j)
	    		    	  				   & (available (rho T j) p))
	    		    	  	  )			    
	        	  		)  
					)
		) 
	)

#----------------------------------------------------------------------------
#-- Send and receive transitions for a nonfaulty actor are IOE

	define nonfaulty->IOE-rcv :=
		(forall x .
			(nonfaulty x) ==>
			(forall m T i .
		        ((inMSet m (mu (config (rho T i)))) 
		        & (ready-to (rho T i) x (receive x m)))  
	            	==>
	            	(
	            		  (exists j . (i N.<= j) 
	            		   			& (= (rho T (S j)) (then (rho T j) (receive x m))) )                  
	            	    |  (forall k .  (i N.< k)
	            	    				==>
	            	    				(exists j .
		            	    				  (k N.<= j) 
		            	    				& (inMSet m (mu (config (rho T j)))) 
		            	    				& (ready-to (rho T j) x (receive x m))
		            	    			)
	            	    	)			              
	            	) 
    		)
     	)	

	define nonfaulty->IOE-snd :=
		(forall x m T i .
				(nonfaulty x) ==>
				(
			        (ready-to (rho T i) x (send x m))  
		            	==>
		            	(
		            		  (exists j . (i N.<= j) 
		            		   			& (= (rho T (S j)) (then (rho T j) (send x m))) )
		            	    |  (forall k .  (i N.< k)
		            	    				==>
		            	    				(exists j .
			            	    				  (k N.<= j) 
			            	    				& (ready-to (rho T j) x (send x m))
			            	    			)
		            	    	) 
		            	) 
	            )	
        )

	define nonfaulty-def-axioms := [ proposer-nonfaulty1
	                            	  proposer-nonfaulty2
	                              	  acceptor-nonfaulty1
	                              	  acceptor-nonfaulty2 
	                              	  nonfaulty->IOE-snd
	                              	  nonfaulty->IOE-rcv
	                            	]

#----------------------------------------------------------------------------
#-- PROPERTIES OF A BALLOT NECESSARY FOR IT TO SUCCESSFULLY GET VOTES 

#-- 

	#-- We can call it the "P1+P2-True" since it requires >= properties
 	#-- If P1+P2-True proposer, then two things: 
 	#--		1. if an acceptor receives 1a message at some indexed position with the ballot,
 	#--        the ballot should be > beta of every acceptor 
 	#-- 	2. if an acceptor receives 2a message at some indexed position with the ballot, 
 	#--        the ballot should be >= beta of every acceptor

 	define P1+P2-True-1a-beta :=                                                                  
 		(forall b . 
 			 	(P1+P2-True b)                                                                    
				 ==> 
				 	(forall p T a i .
				 		(
					 		(
					 			  (acc a) 
						 		& (inMSet (consM p a 1a b nilV) (amu (als (config (rho T i)) a))) 
							)
							==>	((beta (als (config (rho T i)) a)) < b) # In an actual implementation, the beta will be updated after responding to a message
				 		)												# and if a message has been responded to, it will be removed from local state
				 	)													# Therefore, this cannot lead to a fallacy when beta is updated
		)		 	

 	define P1+P2-True-2a-beta :=
 		(forall b . 
 			(P1+P2-True b)
				 ==> 
				 	(forall p T a i v .
				 		(
					 		(
					 			  (acc a) 
						 		& (inMSet (consM p a 2a b v) (amu (als (config (rho T i)) a))) 
							)
							==>	((beta (als (config (rho T i)) a)) =< b)
						) 
				 	)
	 	)

	define P1+P2-True-beta-def-axioms := [ P1+P2-True-1a-beta 
				  					       P1+P2-True-2a-beta 
				  					     ]

#----------------------------------------------------------------------------
#-- CONDITIONS FOR PROVING PROGRESS
#----------------------------------------------------------------------------
 	#-- The following conditions are necessary:
 	#-- In every transaction path, some nonfaulty proposer will eventually be ready to propose a ballot 
 	#-- that satisfies P1+P2 to some quorum of nonfaulty acceptors

 	define progress_conditions :=
	(forall T .
		(exists i p b Q .
					  (prp p) 
					& (P1+P2-True b)
					& (nonfaulty p) 
					& (forall a . (inQ a Q) 
						                ==> 
						                      (nonfaulty a) 
						                    & (ready-to (rho T i) p (send p (consM p a 1a b nilV)))
					  )
		)
	)

}#close module
\end{lstlisting}

\begin{lstlisting}[style=athena-list]
#****************************************************************************
# Helpful reusable methods

#****************************************************************************

extend-module Synod {

  define (make-cond-def-1 neg) := 
    method (premise) 
      match premise {
        (p ==> q) => let {p' := (neg p);
                          goal := (p' | q)}
                       (!by-contradiction goal
                          assume -goal := (~ goal)
                            let {_ := (!dm' -goal);
                                 p := try { (!dn (~ ~ p)) | (!claim p) }}
                              (!absurd (!mp premise p) 
                                       (~ q)))
      }

  define (make-cond-def-2 neg) := 
    method (premise)
      match premise {
         (p | q) => assume p' := (neg p)
                      (!cases premise
                              assume p 
                                (!from-complements q p p')
                              assume q
                                (!claim q))
      }

  define (make-cond-def neg) :=
    method (premise) 
       match premise {
         (_ ==> _) => (!(make-cond-def-1 neg) premise)
       | (_ | _)   => (!(make-cond-def-2 neg) premise)
       }

  define cond-def := (make-cond-def complement)

}# close module Synod 
\end{lstlisting}

\lstinputlisting[style=athena-list]{codes/progress-proof.txt}

\end{appendices}